\documentclass[english,prb,aps,twocolumn,longbibliography,floatfix,showpacs,superscriptaddress]{revtex4-1}
\usepackage[T1]{fontenc}
\usepackage[latin9]{inputenc}
\usepackage{color}
\usepackage{rotating}
\usepackage{bm}
\usepackage{amsmath}
\usepackage{amssymb}
\usepackage{graphicx}

\makeatletter

\providecommand{\tabularnewline}{\\}

\usepackage[colorlinks=true,linkcolor=blue,anchorcolor=blue,citecolor=blue,urlcolor=blue]{hyperref}
\usepackage{lipsum}

\makeatother

\usepackage{babel}
\begin{document}
\renewcommand{\figurename}{Fig.}
\title{Topological and holonomic quantum computation based on second-order topological superconductors}
\author{Song-Bo Zhang}
\email{songbo.zhang@physik.uni-wuerzburg.de}

\affiliation{Institute for Theoretical Physics and Astrophysics$\text{,}$  University of W\"urzburg, D-97074 W\"urzburg, Germany}
\author{W. B. Rui}
\email{wenbin.rui@gmail.com}

\affiliation{Max-Planck-Institute for Solid State Research, Heisenbergstrasse 1, D-70569 Stuttgart, Germany}
\author{Alessio Calzona}
\affiliation{Institute for Theoretical Physics and Astrophysics$\text{,}$  University of W\"urzburg, D-97074 W\"urzburg, Germany}
\author{Sang-Jun Choi}
\affiliation{Institute for Theoretical Physics and Astrophysics$\text{,}$  University of W\"urzburg, D-97074 W\"urzburg, Germany}
\author{Andreas P. Schnyder}
\affiliation{Max-Planck-Institute for Solid State Research, Heisenbergstrasse 1, D-70569 Stuttgart, Germany}
\author{Björn Trauzettel}
\affiliation{Institute for Theoretical Physics and Astrophysics$\text{,}$  University of W\"urzburg, D-97074 W\"urzburg, Germany}
\affiliation{W\"urzburg-Dresden Cluster of Excellence ct.qmat, Germany}
\date{\today}
\begin{abstract}
 Majorana fermions feature non-Abelian exchange statistics and promise fascinating applications in topological quantum computation. Recently, second-order topological superconductors (SOTSs) have been proposed to host Majorana fermions as localized quasiparticles with zero excitation energy, pointing out a new avenue to facilitate topological quantum computation. We provide a minimal model for SOTSs and systematically analyze the features of Majorana zero modes with analytical and numerical methods. We further construct the fundamental fusion principles of zero modes stemming from a single or multiple SOTS islands. Finally, we propose concrete schemes in different setups formed by SOTSs, enabling us to exchange and fuse the zero modes for non-Abelian braiding and holonomic quantum gate operations.
\end{abstract}
\maketitle

\section{Introduction}
Majorana fermions are self-conjugate fermions \citep{Majorana32NC}. They can arise as zero-energy Bogoliubov quasiparticles in condensed
matter \citep{Kitaev01PU,Alicea12PPP,Beenakker13ARCMP,Elliott15RMP},
such as vortex bound states in $p$-wave superconductors \citep{Read00PRB,Ivanov01PRL},
Majorana bound states in Josephson junctions \citep{Fu08PRL,Lutchyn10PRL},
and end states of nanowires with Rashba spin-orbit coupling or of
ferromagnetic atomic chains \citep{Sau10PRL,Oreg10PRL,Choy11PRB,Alicea10PRB,Nadj-Perge13PRB,Nadj-Perge14Science}.
These bound states have zero excitation energy and are commonly coined Majorana zero modes (MZMs).
MZMs can be viewed as one half of ordinary complex fermions and always come in pairs. When more than two MZMs are present, the braiding (exchange) operations on them correspond
to non-Abelian rotations in the ground-state manifold spanned by them.
They can thus serve as basic building blocks for topological quantum computation \citep{Ivanov01PRL,Kitaev03AnPhy,Nayak08RMP,Alicea12PPP,Sarma15npjQI}.
If fusions between MZMs are adiabatically tunable, they can
also be exploited for holonomic quantum gates \citep{Zanardi99PPA,Karzig16PRX}.
Hence, how to nucleate, fuse and braid MZMs in solid-state systems
is one of the main focuses in modern condensed matter physics and quantum
computer science.

\begin{figure}[htp]
\centering

\includegraphics[width=0.9\columnwidth]{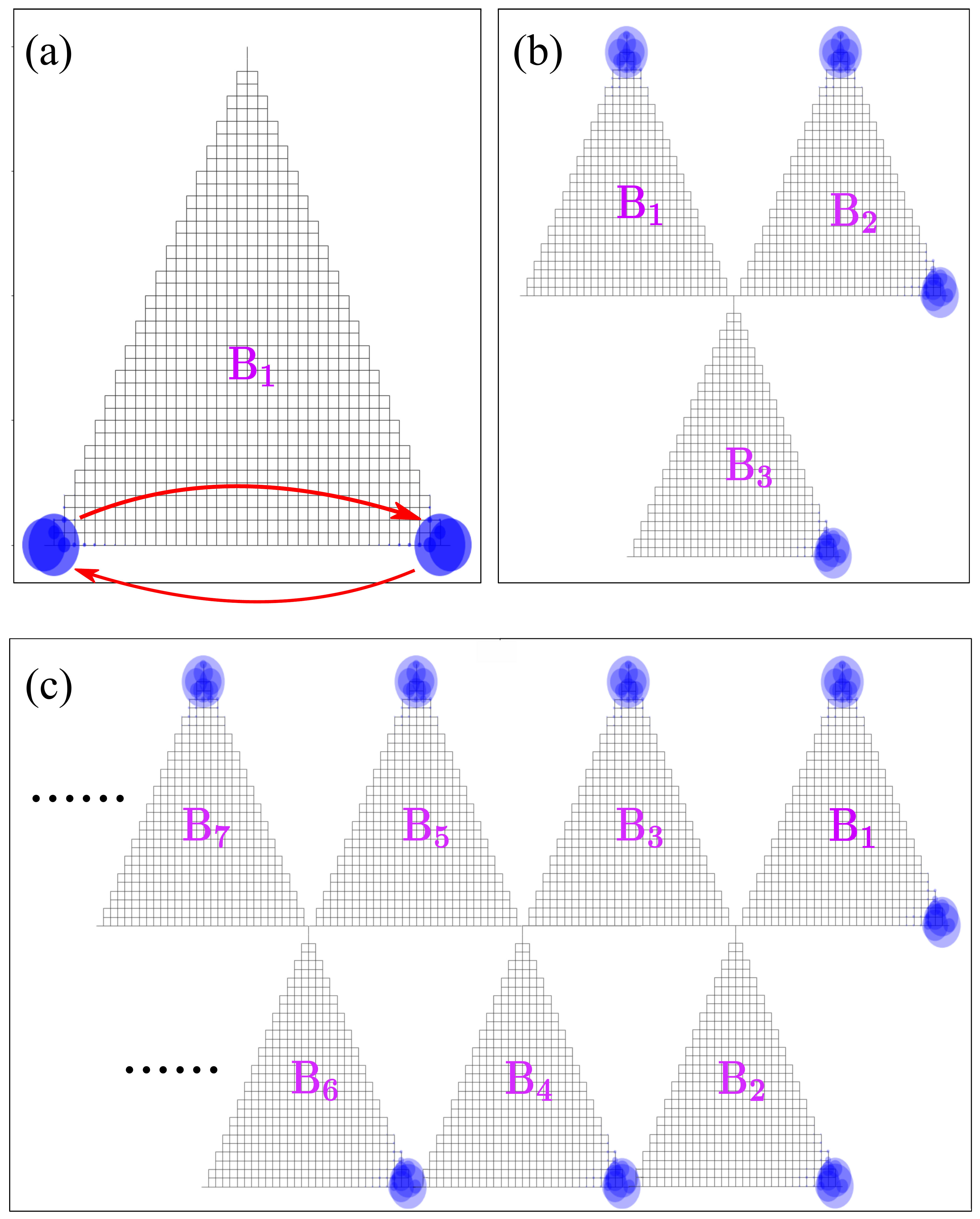}

\caption{Schematics of the SOTS-based setups for braiding (a) two, (b) four, and (c) more MZMs. The setups are scalable in a straightforward way. ${\bf B}_{i}$ with $i=1,2,...$ are
in-plane magnetic fields applied to the triangle islands. The blue
dots indicate the positions of the MZMs.}
\label{fig:main-results}
\end{figure}

Recently, second-order topological superconductors (SOTSs) have been
discovered in various candidate systems and predicted to host localized
MZMs in two dimensions lower than the gapped bulk \citep{Langbehn17PRL,QYWang18PRL,ZBYan18PRL,TLiu18PRB,Geier18PRB, CHHsu18PRL,Volpez19PRL,Wang18PRB,Shapourian18PRB,ZhangSB20PRR,Ghorashi19PRB,Wu19PRB, RXZhang19PRLb,Plekhanov19PRR,Skurativska20PRR,Ahn2019arXiv,Franca19PRB,RXZhang19PRL, Bultinck19PRB,HSu19arXiv,ZBYan19PRL,PanXH19PRL,Bomantara20PRB,Laubscher20PRR,WYJ19arxiv,XXWu19arXiv190510648W,XYZhu18PRB}.
This opens up a new avenue towards Majorana-mediated topological quantum computation. Preliminary attempts have been made in this direction
\citep{XYZhu18PRB,Ezawa19PRB,Pahomi19arXiv}, which are, however,
limited to only two MZMs. A comprehensive study of creating, fusing
and braiding MZMs in SOTSs is lacking. Importantly, the fusion
and braiding of more than two MZMs are essential for a successful
implementation of (topological) quantum gates \citep{Bravyi06PRA,Nayak08RMP}.

In this article, we show that the desired non-Abelian braiding operations
of MZMs as well as a full set of holonomic gates can be achieved
in the SOTS platform. To elucidate this, we provide a minimal
model for SOTSs with inversion symmetry and discuss the features and behaviors of individual
MZMs in a disk geometry, both analytically and numerically. We find
that a finite chemical potential breaks an effective mirror symmetry
and thus gives rise to tunable spin polarizations of the MZMs. Interestingly,
the positions of the MZMs can be controlled by chemical potential
and applied in-plane magnetic field. We systematically analyze
the tunneling interaction between adjacent MZMs stemming from a single or multiple
SOTS islands and identify the fundamental fusion principles between
them. As an illustration of these principles, we demonstrate how to
manipulate the fusion of MZMs in an incomplete disk by tuning
magnetic field and chemical potential.

We put forward a number of setups formed by the SOTSs, as sketched in Fig.\ \ref{fig:main-results}, and present in detail corresponding schemes for braiding the MZMs based on variations of chemical potential, applied magnetic field, and geometry engineering. In a simple triangle setup, we can exchange two MZMs located at the vertices by rotating the in-plane magnetic field applied to the setup. In a trijunction setup constructed by three triangle islands, we are able to exchange any two of four MZMs by purely rotating the magnetic fields or by tuning both the directions and strength of the magnetic fields.
Our theory can be extended to a ladder structure which is formed by elementary triangle islands and hosts any number of MZM pairs for braiding performance, see Fig.\ \ref{fig:main-results}(c). Hence, our proposal is scalable in a straightforward way. Moreover, we propose a shamrock-like trijunction setup constituted by three incomplete disks. This trijunction hosts three MZMs that fuse exclusively with a fourth
one. The fusion strengths are smoothly adjustable providing a feasible platform for holonomic quantum gates. Our schemes could be tested, for instance, in quantum spin Hall insulators (QSHIs) in proximity to conventional superconductors or in monolayer
$\text{FeTe}{}_{1-x}\text{Se}_{x}$. Furthermore,
they reveal the innovative idea of geometry engineering to braid and
fuse MZMs based on SOTSs.

The article is organized as follows. In Sec.\ \ref{sec:Model-Hamiltonian-and}, we introduce the minimal model for SOTSs, derive an effective boundary Hamiltonian and the wavefunctions and spin polarizations of MZMs. Next, we analyze the fusion properties of the MZMs in Sec.\ \ref{sec:Fusion-rules-of}. We proceed to describe the setups and schemes for braiding two or more MZMs, and discuss the important relevant physics in Sec.\ \ref{sec:braiding}. We devote Sec.\ \ref{sec:holonomic} to our proposal of holonomic gates. Finally, we discuss the experimental implementation and measurement, and summarize the results in Sec.\ \ref{sec:discussion}.

\section{Model Hamiltonians and Majorana zero modes} \label{sec:Model-Hamiltonian-and}
\subsection{Effective boundary Hamiltonian}

We start with considering simple two-dimensional SOTSs realized, for instance, in QSHIs in presence of superconductivity and moderate in-plane magnetic fields \citep{XYZhu18PRB,XXWu19arXiv190510648W,PanXH19PRL,WYJ19arxiv}. The minimal model for the SOTSs can be written as $\mathcal{H}=\mathcal{H}_{0}+\mathcal{H}_{\text{ex}}$
with
\begin{eqnarray}
\mathcal{H}_{0} & = & m({\bf k})\tau_{z}\sigma_{z}+v\sin k_{x}s_{z}\sigma_{x}+v\sin k_{y}\tau_{z}\sigma_{y}-\mu\tau_{z},\nonumber \\
\mathcal{H}_{\text{ex}} & = & \Delta({\bf k})\tau_{y}s_{y}+g\mu_{B}B(\cos\theta\tau_{z}s_{x}+\sin\theta s_{y}\sigma_{z}),\label{eq:minimal-model}
\end{eqnarray}
where $m({\bf k})=m_{0}-2m(2-\cos k_{x}-\cos k_{y})$. Without loss
of generality, we assume positive parameters $v$, $m$ and $m_{0}$,
and set the lattice constant $a$ to unity. ${\bf s}$, $\bm{{\bf \sigma}}$
and $\bm{\tau}$ are Pauli matrices acting on spin, orbital, and Nambu
spaces, respectively. $\mu$ is the chemical potential. The pairing
interaction can be $s$-wave or $s_{\pm}$-wave for our purposes. However, the exact
form of $\Delta({\bf k})$ is not important for our main results.
We take it as a constant $\Delta({\bf k})=\Delta_{0}>0$ for simplicity.
${\bf B}\equiv B(\cos\theta,\sin\theta)$ is the magnetic field with
strength $B$ and direction $\theta$. The effective g-factors
of the two orbitals are the same in the $x$ direction and opposite
in the $y$ direction. This model applies to an inverted HgTe quantum well \citep{Bernevig06science,Durnev16PRB} with
proximity-induced superconductivity, or a monolayer $\text{FeTe}{}_{1-x}\text{Se}_{x}$ \citep{XXWu16PRB,XXWu19arXiv190510648W} under in-plane magnetic fields.

The minimal model\ (\ref{eq:minimal-model}), in general, has only one
crystalline symmetry, inversion symmetry. The existence of MZMs
is not restricted to any specific geometry \citep{Khalaf18PRB}. Moreover,
the MZMs appear as low-energy quasiparticles at open boundaries. Thus,
to explore the MZMs, we start from the $k\cdot p$ limit of the model\ (\ref{eq:minimal-model})
and consider the SOTS in a large disk geometry. In the absence of
${\bf B}$, this low-energy model respects an emergent in-plane rotational
symmetry. We first derive the boundary states of $\mathcal{H}_{0}$
which is decoupled into four blocks representing Dirac Hamiltonians.
In the disk, the angular momentum $\nu$ is a good quantum number.
It is thus convenient to work in polar coordinates: $r=\sqrt{x^{2}+y^{2}}$
and $\varphi=\arctan(y/x)$. The application of ${\bf B}$ will break
this symmetry. However, in a large disk (with radius $R\gg m/v$),
we can approximate a small segment of the boundary at an arbitrary angle
$\varphi$ as a straight line. Define an effective coordinate $s\equiv R\varphi$
along the segment and treat the corresponding momentum $p_{\nu}\equiv\nu/R$ as a quasi-good quantum number. Then, the energy bands of the boundary
states of the four blocks can be derived as
\begin{eqnarray}
E_{e,\uparrow/\downarrow}(p_{\nu}) & = & \mp vp_{\nu}-\mu,\nonumber \\
E_{h,\uparrow/\downarrow}(p_{\nu}) & = & \mp vp_{\nu}+\mu,
\end{eqnarray}
respectively. The boundary states are helical with velocity $v$.
Correspondingly, the wavefunctions can be written as
\begin{equation}
\Psi_{e,\uparrow,p_{\nu}}=e^{ip_{\nu}s}K(r)(1,-ie^{i\varphi},0,0,0,0,0,0)^{T},\label{eq:basis}
\end{equation}
$\Psi_{e,\downarrow,p_{\nu}}=is_{y}\Psi_{e,\uparrow,-p_{\nu}}^{*}$and
$\Psi_{h,\uparrow/\downarrow,p_{\nu}}=\tau_{x}\Psi_{e,\uparrow/\downarrow,-p_{\nu}}^{*}$,
where $K(r)=\mathcal{N}e^{ip_{\nu}s}[e^{\lambda_{1}(r-R)}-e^{\lambda_{2}(r-R)}]$,
$\lambda_{1/2}=v/2m\pm[(v/2m)^{2}-m_{0}/m+p_{\nu}^{2}]^{1/2}$ and the prefactor
$\mathcal{N}$ takes care of normalization. We provide more details
of this derivation in App. \ref{sec:Edge-states-in}.

With $(\Psi_{e,\uparrow,p_{\nu}},\Psi_{e,\downarrow,p_{\nu}},\Psi_{h,\uparrow,p_{\nu}},\Psi_{h,\downarrow,p_{\nu}})$
as basis, we project the full Hamiltonian onto the boundary
states. The resulting effective Hamiltonian on the boundary can be
written as
\begin{eqnarray}
\mathcal{\widetilde{H}} & = & \begin{pmatrix}-vp_{\nu}-\mu & ie^{-i\varphi}\widetilde{B} & 0 & -\Delta_{0}\\
-ie^{i\varphi}\widetilde{B} & vp_{\nu}-\mu & \Delta_{0} & 0\\
0 & \Delta_{0} & -vp_{\nu}+\mu & ie^{i\varphi}\widetilde{B}\\
-\Delta_{0} & 0 & -ie^{-i\varphi}\widetilde{B} & vp_{\nu}+\mu
\end{pmatrix},\label{eq:boundary-Hamiltonian}
\end{eqnarray}
where $\widetilde{B}=B\sin(\varphi-\theta)$ and $g\mu_{B}=1$ has been chosen for convenience. The effective pairing potential $\Delta_{0}$
felt by the boundary states is constant and independent of the boundary orientation. It couples electrons to holes with opposite spins and
preserves time-reversal and in-plane rotational symmetries. Thus,
the energy spectrum is fully gapped in the absence of $B$. This implies
that our choice of the pairing interaction alone cannot lead to a second-order topological
phenomenon. The scenario becomes different when we turn
on $B$. The magnetic field couples states with opposite spins and
breaks the rotational symmetry. The effective magnetic field $\widetilde{B}$
depends substantially on the angular position $\varphi$ (or equivalently,
the boundary orientation, which is parallel to the azimuthal direction
at $\varphi$). Consequently, interesting physics including controllable
MZMs arise, which we discuss in detail below.

\subsection{MZMs and their positions}
When $\mu=0$, through a unitary transformation $U(\varphi)=$ $e^{-i\pi\tau_{x}/4}e^{i\varphi\tau_{z}s_{z}/2}$,
where ${\bf s}$ and $\bm{\tau}$ are Pauli matrices acting on spin
and Nambu spaces of boundary states, respectively, Eq.\,(\ref{eq:boundary-Hamiltonian})
can be brought into block-diagonal form
\begin{equation}
U(\varphi)\mathcal{\widetilde{H}}(\varphi)U^{-1}(\varphi)=h_{\text{u}}\oplus h_{\text{d}},\label{eq:transform-H}
\end{equation}
where $h_{\text{u}/\text{d}}=-vp_{\nu}s_{z}+(\widetilde{B}\pm\Delta_{0})s_{y}$.
The two blocks $h_{\text{u/d}}$ are essentially one-dimensional Dirac
Hamiltonians with the masses given by $\widetilde{B}\pm\Delta_{0}$,
respectively. They can be connected not only by inversion $\mathcal{P}$
but also by an effective mirror symmetry $\mathcal{M}$ with the mirror
line in the field direction. Note that $\mathcal{P}$ and $\mathcal{M}$
act nonlocally on the boundary model, i.e., $\widetilde{\mathcal{P}}\mathcal{\widetilde{H}}(p_{\nu},\varphi)\widetilde{\mathcal{P}}^{-1}=\mathcal{\widetilde{H}}(p_{\nu},\varphi+\pi)$
and $\widetilde{\mathcal{M}}\mathcal{\widetilde{H}}(p_{\nu},\varphi-\theta)\widetilde{\mathcal{M}}^{-1}=\mathcal{\widetilde{H}}(-p_{\nu},\theta-\varphi)$,
where the overhead tilde ($\widetilde{\ \ }$) indicates boundary-state
space. For $h_{\text{d}}$, the inclusion of $B>\Delta_{0}$ changes
the sign of the Dirac mass at $\varphi_{1}=\theta+\text{\ensuremath{\arcsin}}(\Delta_{0}/B)$ and $\varphi_{2}=\theta-\text{\ensuremath{\arcsin}}(\Delta_{0}/B)+\pi$.
Thus, two localized Majorana states with zero energy appear, which
we label as $\gamma_{1}$ and $\gamma_{2}$. If inversion symmetry
is present, then another pair of MZMs, labeled as $\gamma_{3}$ and
$\gamma_{4}$, can be found from $h_{\text{u}}$. They are located
at the positions different from $\varphi_{1}$ and $\varphi_{2}$
by an angle $\pi$.

\begin{figure}[htp]
\centering \includegraphics[width=1\columnwidth]{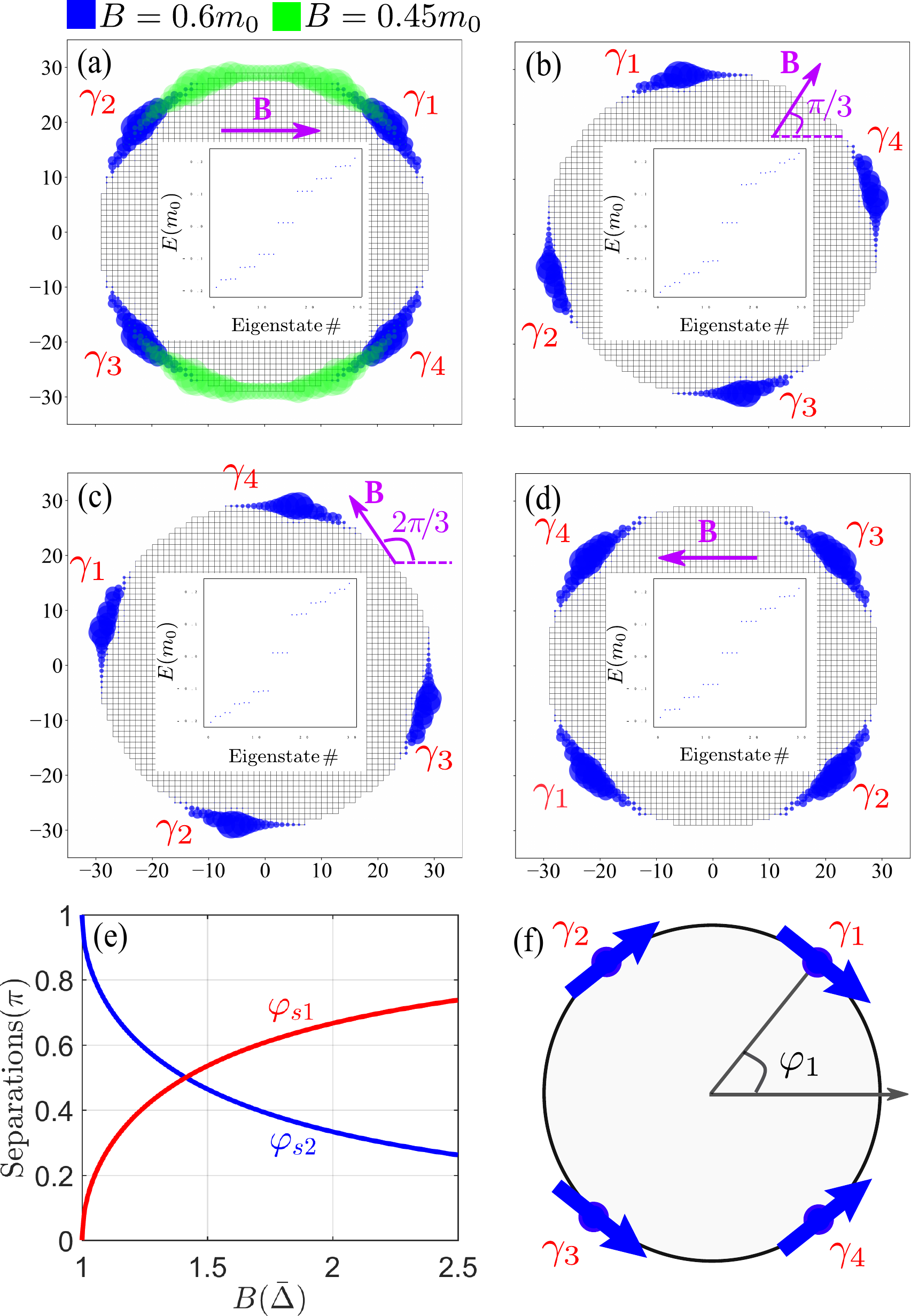}
\caption{Majorana zero modes in a disk geometry. (a)-(d) The blue
color indicates the positions of the MZMs for magnetic fields in different
directions, (a) $\theta=0$, (b) $\pi/3$, (c) $2\pi/3$ and (d) $\pi$,
respectively. By rotating the magnetic field ${\bf B}$, the MZMs
circle around the disk boundary. Inserts are the corresponding energy
spectra. Four MZMs are protected by an energy gap from excited modes.
The positions of the MZMs depend also on the strength of magnetic
field $B$ and chemical potential $\mu$. (e) Separations $\varphi_{1s}$
and $\varphi_{2s}$ as functions of $B$ (in units of $\bar{\Delta}$).
(f) Schematic of the spin polarizations of the MZMs at a finite chemical
potential $\mu$. Here, we choose the parameters as $\mu=0.05m_{0}$, $\Delta_{0}=0.4m_{0}$, $B=0.6m_{0}$, $v=m=m_{0}=1$ and $N_{r}=30a$. $a$ is the lattice constant. For the green color in (a), we use $B=0.45m_{0}$.}
\label{fig:MajoranaModes-complete-disk}
\end{figure}

\begin{table*}[t]
\begin{tabular}{|c||c|c|c|c|}
\hline
 & $\gamma_{1}$ & $\gamma_{2}$ & $\gamma_{3}$ & $\gamma_{4}$\tabularnewline
\hline
\hline
Position $\varphi_{i}$\textcolor{white}{$\dfrac{d}{}$} & $\theta+\text{arc\ensuremath{\sin}}(\bar{\Delta}/B)$ & $\theta-\text{arc\ensuremath{\sin}}(\bar{\Delta}/B)+\pi$ & $\theta+\text{arc\ensuremath{\sin}}(\bar{\Delta}/B)+\pi$ & $\theta-\text{arc\ensuremath{\sin}}(\bar{\Delta}/B)$\tabularnewline
\hline
$\begin{array}{c}
\text{Wavefunction}\\
\Psi_{i}
\end{array}$ & $\mathcal{F}_1\begin{pmatrix}e^{-i(\varphi_{1}-\vartheta+\pi/2)/2}\\
-e^{i(\varphi_{1}+\vartheta+\pi/2)/2}\\
e^{i(\varphi_{1}-\vartheta-\pi/2)/2}\\
e^{-i(\varphi_{1}+\vartheta-\pi/2)/2}\\
e^{i(\varphi_{1}-\vartheta+\pi/2)/2}\\
-e^{-i(\varphi_{1}+\vartheta+\pi/2)/2}\\
e^{-i(\varphi_{1}-\vartheta-\pi/2)/2}\\
e^{i(\varphi_{1}+\vartheta-\pi/2)/2}
\end{pmatrix}$ & $\mathcal{F}_{2}\begin{pmatrix}-e^{-i(\varphi_{2}+\vartheta+\pi/2)/2}\\
e^{i(\varphi_{2}-\vartheta+\pi/2)/2}\\
e^{i(\varphi_{2}+\vartheta-\pi/2)/2}\\
e^{-i(\varphi_{2}-\vartheta-\pi/2)/2}\\
-e^{i(\varphi_{2}+\vartheta+\pi/2)/2}\\
e^{-i(\varphi_{2}-\vartheta+\pi/2)/2}\\
e^{-i(\varphi_{2}+\vartheta-\pi/2)/2}\\
e^{i(\varphi_{2}-\vartheta-\pi/2)/2}
\end{pmatrix}$ & $\mathcal{F}_{3}\begin{pmatrix}e^{-i(\varphi_{3}-\vartheta-\pi/2)/2}\\
e^{i(\varphi_{3}+\vartheta-\pi/2)/2}\\
-e^{i(\varphi_{3}-\vartheta+\pi/2)/2}\\
e^{-i(\varphi_{3}+\vartheta+\pi/2)/2}\\
e^{i(\varphi_{3}-\vartheta-\pi/2)/2}\\
e^{-i(\varphi_{3}+\vartheta-\pi/2)/2}\\
-e^{-i(\varphi_{3}-\vartheta+\pi/2)/2}\\
e^{i(\varphi_{3}+\vartheta+\pi/2)/2}
\end{pmatrix}$ & $\mathcal{F}_{4}\begin{pmatrix}e^{-i(\varphi_{4}+\vartheta-\pi/2)/2}\\
e^{i(\varphi_{4}-\vartheta-\pi/2)/2}\\
e^{i(\varphi_{4}+\vartheta+\pi/2)/2}\\
-e^{-i(\varphi_{4}-\vartheta+\pi/2)/2}\\
e^{i(\varphi_{4}+\vartheta-\pi/2)/2}\\
e^{-i(\varphi_{4}-\vartheta-\pi/2)/2}\\
e^{-i(\varphi_{4}+\vartheta+\pi/2)/2}\\
-e^{i(\varphi_{4}-\vartheta+\pi/2)/2}
\end{pmatrix}$\tabularnewline
\hline
\multicolumn{1}{|c||}{Polarization\textcolor{white}{$\dfrac{l}{l}$}${\bf S}^{(e/h)}$} & $\pm S_{0}(\sin\varphi_{1},-\cos\varphi_{1},0)$ & $\pm S_{0}(\sin\varphi_{2},-\cos\varphi_{2},0)$ & $\pm S_{0}(-\sin\varphi_{3},\cos\varphi_{3},0)$ & $\pm S_{0}(-\sin\varphi_{4},\cos\varphi_{4},0)$\tabularnewline
\hline
\end{tabular}
\caption{Wavefunctions and spin polarizations of four MZMs on a disk boundary. The wavefunctions are written in the basis of the minimal model\ (\ref{eq:minimal-model}). The prefactors $\mathcal{F}_{i}$ account for the spatial distribution near the positions ($R$, $\varphi_i$) and for normalization (see App.\ \ref{sec:Wavefunction-and-polarization} for more details). We define $\vartheta\equiv \text{arctan}(\mu/\Delta_{0})$. The magnitude of the spin polarizations ${\bf S}^{(e/h)}\equiv(S_{x}^{(e/h)},S_{y}^{(e/h)},S_{z}^{(e/h)})$ is $S_{0}=\hbar\mu/(4\bar{\Delta})$. The subscripts $(e/h)$ stand for the electrons/holes. The spin polarizations of the electron and hole parts are opposite, $\text{\ensuremath{\bm{{\bf S}}_{\gamma_{i}}^{(h)}}}=-\text{\ensuremath{\bm{{\bf S}}_{\gamma_{i}}^{(e)}}}$.}
\label{table1}
\end{table*}

When $\mu\neq0$, the transformed Hamiltonian\ (\ref{eq:transform-H}) is no longer block diagonal. Nevertheless, the four bands of the Hamiltonian
can still be analytically derived:
\begin{equation}
E=\pm\sqrt{\widetilde{B}^{2}+A^{2}p_{\nu}^{2}+\bar{\Delta}^{2}\pm2\sqrt{\widetilde{B}^{2}\bar{\Delta}^{2}+A^{2}\mu^{2}p_{\nu}^{2}}},\label{eq:energy-band}
\end{equation}
where $\bar{\Delta}=\sqrt{\Delta_{0}^{2}+\mu^{2}}$. Increasing $|\widetilde{B}|$
from $0$ to a value larger than $\bar{\Delta}$, we observe band
inversions happening at $p_{\nu}=0$. Suppose that $B$ is sufficiently
large, $B^{2}>\bar{\Delta}^{2}$, then, due to the oscillatory behavior
of $\widetilde{B}$, the band order changes when moving along the
disk boundary. This indicates the appearance of MZMs. The positions
of the MZMs are determined by the closing points of the bands.
Thus, the positions $\varphi_{i}$ of MZMs $\gamma_{i}$ (with
$i\in\{1,2,3,4\}$) are generally given by
\begin{eqnarray}
\varphi_{1/4} & = & \text{\ensuremath{\theta}\ensuremath{\ensuremath{\pm}}\ensuremath{\arcsin}}(\bar{\Delta}/B),\nonumber \\
\varphi_{2/3} & = & \theta\mp\text{\ensuremath{\arcsin}}(\bar{\Delta}/B)+\pi,\label{eq:positions}
\end{eqnarray}
where we choose the convention $0\leqslant \text{\ensuremath{\arcsin}}(\bar{\Delta}/B)<\pi/2$. Similar to the MZMs in the $\mu=0$
limit, $\gamma_{1}$ ($\gamma_{2}$) and $\gamma_{3}$ ($\gamma_{4}$) are always separated by an angle $\pi$. They are related to each
other by inversion symmetry. In contrast, the separations between
the neighboring MZMs $\gamma_{1}$ ($\gamma_{2}$) and $\gamma_{4}$
($\gamma_{3}$) and $\gamma_{1}$ ($\gamma_{3}$) and $\gamma_{2}$
($\gamma_{4}$) are respectively given by
\begin{equation}
\varphi_{s1}=2\arcsin(\bar{\Delta}/B) \text{ and }\varphi_{s2}=\pi-\varphi_{s1}.\label{eq:separations}
\end{equation}
These separations are independent of the field direction $\theta$. However, they are controllable by the chemical potential $\mu$ and field strength $B$, via the ratio $\bar{\Delta}/B$. Increasing $B$ or decreasing $\mu$, $\varphi_{s1}$ is monotonically increased whereas $\varphi_{2s}$ is decreased, as shown in Fig.\ \ref{fig:MajoranaModes-complete-disk}(e).

Interestingly, the positions of $\gamma_{i}$ depend not only on the ratio $\bar{\Delta}/B$ but also on the direction of the magnetic field, according to Eq.\ (\ref{eq:positions}). When rotating ${\bf B}$, the MZMs $\gamma_{i}$ move around the disk boundary. To confirm these features,
we employ the tight-binding model (\ref{eq:minimal-model}) and define the disk by $i_{x}^{2}+i_{y}^{2}\leq N_{r}^{2},$ where $N_{r}$ is
the radius of the disk and $i_{x},i_{y}$ label the lattice sites in the $x$ and $y$ directions, respectively. We use the Kwant package \citep{Groth14Kwant} to plot the wavefunctions.  For a large magnetic field $B>\bar{\Delta}$ applied in the $x$ direction, we clearly identify
four MZMs centered at the symmetric positions obeying $\varphi_{1}=-\varphi_{4}$ and $\varphi_{2/3}=\varphi_{4/1}+\pi$, see Fig.\ \ref{fig:MajoranaModes-complete-disk}(a). When rotating ${\bf B}$ from $x$ to $-x$ direction, all the MZMs are moving anticlockwise. However, their separations are unchanged, see Fig.\ \ref{fig:MajoranaModes-complete-disk}(a-d). These observations perfectly agree with our analytical results. Moreover, we find that although the positions of the MZMs are quite different for different $\theta$, the energy spectra of the disks are almost the same. There is always an energy gap protecting the MZMs from excited modes, as long as the MZMs are well separated.

\subsection{Wavefunctions and spin polarizations of MZMs \label{sec:wavefunction}}

With the help of the boundary Hamiltonian\,(\ref{eq:boundary-Hamiltonian}), the wavefunctions of the MZMs $\gamma_{i}$ (with $i\in\{1,2,3,4\}$)
can be analytically derived. We summarize the results in Table\ \ref{table1} and provide the detailed derivations in App. \ref{sec:Wavefunction-and-polarization}.
Several important features can be observed from these wavefunctions. First, the MZMs decay exponentially away from the corresponding centers $\varphi_{i}$, as indicated by the prefactors $\mathcal{F}_i$. Second, up to an overall phase (in the prefactor $\mathcal{F}_i$), the wavefunction of any MZM can be written in a self-adjoint form, $\Psi_{i}=\tau_{x}\Psi_{i}^{*}$, which verifies the basic Majorana property of MZMs. Third, recall that $\gamma_{1}$ ($\gamma_{2}$) and $\gamma_{3}$ ($\gamma_{4}$) are separated by an angle $\pi$. We can find that $\Psi_{3}=\mathcal{P}\Psi_{1}$ and $\Psi_{4}=\mathcal{P}\Psi_{2}$, where $\mathcal{P}=\sigma_{z}$ is the inversion symmetry operator.
If we compare them at the same $\varphi$, then we have instead $\Psi_{3}= \bar{\mathcal{P}} \Psi_{1}$ and $\Psi_{4}=\bar{\mathcal{P}}\Psi_{2}$, where $\bar{\mathcal{P}}\equiv \mathcal{T}_{\pi} \mathcal{P}=\tau_zs_z$ and $\mathcal{T}_{\pi}=e^{i\pi/2}e^{-i\pi\tau_zs_z\sigma_z/2}$ is the operator that shifts the angle $\varphi$ by $\pi$.
These relations reflect the fact that $\gamma_{1}$ ($\gamma_{2}$) is transformed to $\gamma_{3}$ ($\gamma_{4}$) by inversion symmetry, as mentioned before.
Finally, each MZM acquires a quantized Berry phase $\pi$ when it moves adiabatically around the disk boundary. This $\pi$ Berry phase results from the spinor form of the wavefunctions protected by particle-hole symmetry.

\begin{figure*}[ht].
\centering
\includegraphics[width=1\textwidth]{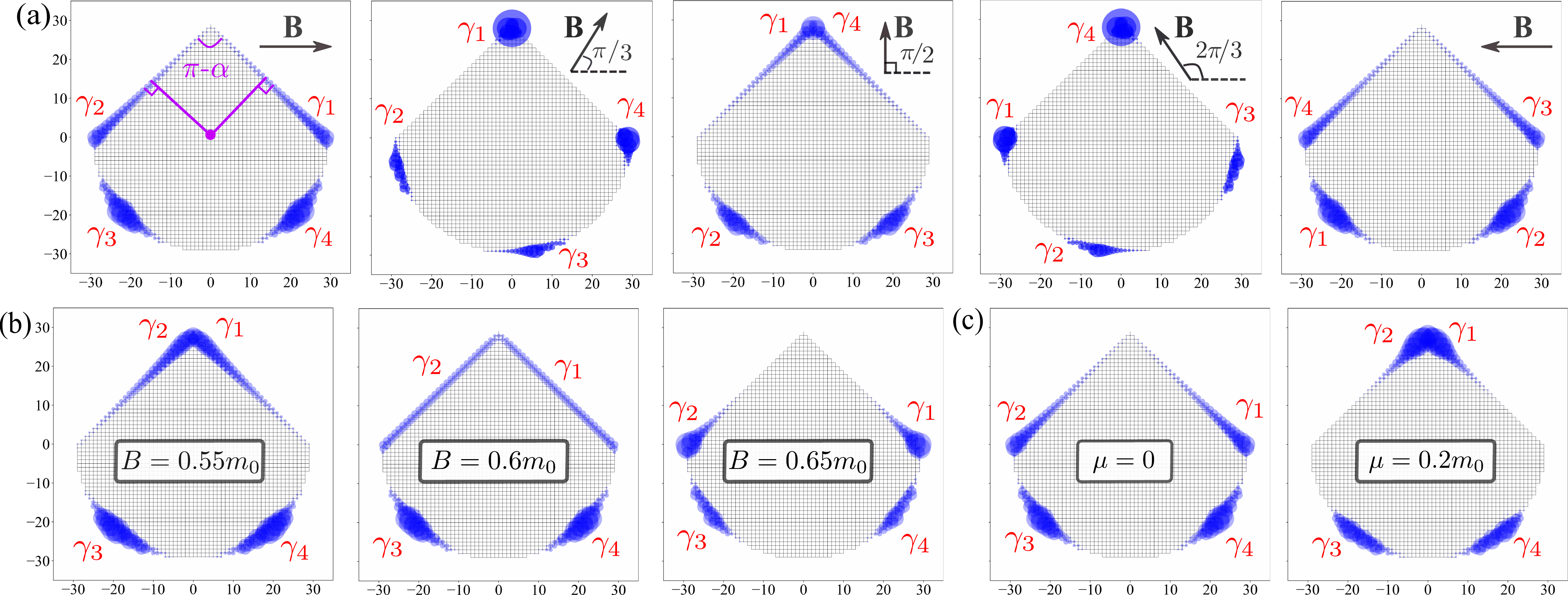}\caption{Manipulation of the fusion of MZMs in an incomplete disk. (a) Control of the fusion by tuning the magnetic field direction $\theta$. At the field direction $\theta=\theta_{c}\in\{0,\pi/4,\pi/2,3\pi/4\}$ modulo $\pi$, two adjacent MZMs collide into the sharp vertex and fuse. Tuning $\theta$ away from $\theta_{c}$ splits the two MZMs from the vertex and weakens their fusion. (b) Control of the fusion by tuning $B$. The MZMs $\gamma_{1}$ and $\gamma_{2}$ are well separated for $B>B_{c1}$, whereas they collide and fuse (at the vertex) for $B<B_{c1}$, where $B_{c1}=\bar{\Delta}/\sin(\alpha/2)$. (c) Control of the fusion by tuning $\mu$. $\gamma_{1}$ and $\gamma_{2}$ are separated for $|\mu|<\mu_{c1}$ with $\mu_{c1}=[B^{2}\sin^{2}(\alpha/2)-\Delta_{0}^{2}]^{1/2}$, whereas they collide and fuse for $|\mu|>\mu_{c1}$. $\mu=0.05m_{0}$ in (a) and $0$ in (b), $B=0.6m_{0}$ and $\theta=0$ in (a,c), and, for all panels, all other parameters are the same as those in Fig.\ \ref{fig:MajoranaModes-complete-disk}}
\label{fig:fusion}
\end{figure*}

When $\mu=0$, the disk system respects the effective mirror symmetry, as mentioned
before. Thus, the MZMs are spinless. However, at finite $\mu\neq0$, the mirror symmetry is no longer preserved. Consequently, the MZMs become spin polarized in the $x$-$y$ plane, see Table\ \ref{table1}. The magnitudes of the spin polarizations are given by $\hbar\mu/(4\bar{\Delta})$. In the limit $|\mu|\gg\Delta_{0}$, the MZMs are fully spin polarized. The directions always point parallel or anti-parallel to the boundary orientation,
as illustrated in Fig.\ \ref{fig:MajoranaModes-complete-disk}(f). Thus, when the MZMs move around the disk boundary, their spin polarizations rotate correspondingly. Since $\gamma_{1}$ and $\gamma_{3}$ are separated by $\pi$, they have the same polarization. Similarly, $\gamma_{2}$ and $\gamma_{4}$
have also an identical polarization.

\section{Fusion between MZMs} \label{sec:Fusion-rules-of}
We proceed to discuss the fusion between
the MZMs. Let us first consider the disk geometry with vanishing $\mu$.
Two adjacent MZMs can be brought close to each other by tuning $\mu$
or $B$, according to Eq.\ (\ref{eq:separations}). If the two MZMs
belong to the same block of Eq.\ (\ref{eq:transform-H}), e.g., $\gamma_{1}$
and $\gamma_{2}$ (or $\gamma_{3}$ and $\gamma_{4}$), then they
fuse with each other and acquire a hybridization energy which depends
exponentially on their distance in space. In contrast, since $h_{\text{u}}$
and $h_{\text{d}}$ do not interact with each other due to the effective
mirror symmetry, $\gamma_{1/2}$ stemming from $h_{\text{u}}$ cannot
fuse with $\gamma_{3/4}$ which stem from the other block $h_{\text{d}}$,
even if they sit at the same position and have strong overlap in wavefunction.
This feature can be exploited for holonomic quantum computation, which we will discuss later.

\begin{table}[b]
\begin{tabular}{|r@{\extracolsep{0pt}.}l||r@{\extracolsep{0pt}.}l|r@{\extracolsep{0pt}.}l|r@{\extracolsep{0pt}.}l|r@{\extracolsep{0pt}.}l|}
\hline
\multicolumn{2}{|c||}{\begin{turn}{90}
\end{turn}} & \multicolumn{2}{c|}{\textcolor{white}{$ $}$\gamma_{1}$\textcolor{white}{$\dfrac{}{}$}} & \multicolumn{2}{c|}{$\gamma_{2}$} & \multicolumn{2}{c|}{$\gamma_{3}$} & \multicolumn{2}{c|}{$\gamma_{4}$\ \ }\tabularnewline
\hline
\hline
\multicolumn{2}{|c||}{\ $\gamma_{1}'$\ } & \multicolumn{2}{c|}{\textcolor{white}{$\dfrac{l}{}$} {0}\textcolor{white}{$\dfrac{l}{}$}} & \multicolumn{2}{c|}{$\cos\vartheta\cos\Phi$} & \multicolumn{2}{c|}{$\sin\Phi$} & \multicolumn{2}{c|}{$\sin\vartheta\cos\Phi$}\tabularnewline
\hline
\multicolumn{2}{|c||}{$\gamma_{2}'$} & \multicolumn{2}{c|}{$\cos\vartheta\cos\Phi$} & \multicolumn{2}{c|}{\textcolor{white}{$\dfrac{l}{}$}{0}\textcolor{white}{$\dfrac{l}{}$}} & \multicolumn{2}{c|}{$\sin\vartheta\cos\Phi$} & \multicolumn{2}{c|}{$\sin\Phi$}\tabularnewline
\hline
\multicolumn{2}{|c||}{$\gamma_{3}'$} & \multicolumn{2}{c|}{$\sin\Phi$} & \multicolumn{2}{c|}{$\sin\vartheta\cos\Phi$} & \multicolumn{2}{c|}{\textcolor{white}{$\dfrac{l}{}$}{0}\textcolor{white}{$\dfrac{l}{}$}} & \multicolumn{2}{c|}{$\cos\vartheta\cos\Phi$}\tabularnewline
\hline
\multicolumn{2}{|c||}{$\gamma_{4}'$} & \multicolumn{2}{c|}{$\sin\vartheta\cos\Phi$} & \multicolumn{2}{c|}{$\sin\Phi$} & \multicolumn{2}{c|}{$\cos\vartheta\cos\Phi$} & \multicolumn{2}{c|}{\textcolor{white}{$\dfrac{l}{}$}{0}{\textcolor{white}{$\dfrac{l}{}$}}}\tabularnewline
\hline
\end{tabular}

\label{tab:fusion-rule}

\caption{Fusion of adjacent MZMs located in two connected SOTS islands. The table shows the dependence of fusion strength on the chemical potential, magnetic field and pairing phase difference. $2\Phi$ is the pairing phase difference between the two islands, and the same $\mu$ is assumed in all islands for simplicity. The results for the case of a single island can be  obtained by taking $\gamma_{i}'=\gamma_{i}$ and $\Phi=0$. More details of the table are given in App. \ref{sec:Fusion-of-Majorana}}

\label{TableII}
\end{table}

A finite $\mu$, however, couples the two blocks. Therefore, the fusion between $\gamma_{1}$ and $\gamma_{4}$ ($\gamma_{2}$ and $\gamma_{3}$) is allowed. In the SOTSs, the fusion of MZMs is realized by the hopping interaction, depending on the momentum-orbital coupling. According to Eq.\ (\ref{eq:minimal-model}), the corresponding operators in the $x$ and $y$ directions can be found as $\hat{T}_{x}=ivs_{z}\sigma_{x}/2+m\tau_z\sigma_z$ and $\hat{T}_{y}=iv\tau_{z}\sigma_{y}/2+m\tau_z\sigma_z$,
respectively. Thus, the fusion strength between two MZMs, say $\gamma_{i}$ and $\gamma_{j}$ (with $i,j\in\{1,2,3,4\}$), can
be calculated as $|\langle\Psi_{i}|(\hat{T}_{x}+\hat{T}_{y})|\Psi_{j}\rangle|$. We summarize the results in Table\ \ref{TableII}. In a homogeneous system, the fusion between $\gamma_{1/2}$ ($\gamma_{3/4}$) is proportional to $\cos\vartheta$, while the one between $\gamma_{1/4}$ ($\gamma_{2/3}$) is linear in $\text{sin}\vartheta$, where $\vartheta=\arctan(\mu/\Delta_{0})$.
The proportionality is determined by the overlap of the wavefunctions
of the two involved MZMs. In contrast, in a single SOTS island, $\gamma_{1}$ and $\gamma_{3}$ ($\gamma_{2}$ and $\gamma_{4}$) are well separated in space. Moreover, they are related by inversion symmetry. Notice that $\hat{T}_{x}$ and $\hat{T}_{y}$ anti-commute with the inversion
operator $\bar{\mathcal{P}}$. The fusion between $\gamma_{1/3}$ ($\gamma_{2/4}$)
is prohibited even in the presence of finite $\mu$. These results
are generic and also apply to the case where the two MZMs belong to
different but connected SOTS islands with the same pairing phase. However,
if two islands have a pairing phase difference $2\Phi\neq2n\pi$ with
$n\in\{0,\pm1,...\}$, then $\gamma_{1}$ ($\gamma_{2}$) stemming from one
island can also fuse with $\gamma_{3}$ ($\gamma_{4}$) from the other
island, see Table\ \ref{TableII}. The fusion induced by a phase
difference can be used to realize the braiding of more than two
MZMs, which we demonstrate in Sec.\ \ref{sec:braiding} below.

In a full disk, all angular positions are available. Thus, the four MZMs related by inversion symmetry appear or disappear simultaneously. To better understand the fusion behavior, it is instructive to consider a geometry that breaks inversion symmetry. For concreteness, we consider in the following an incomplete disk made by cutting off two symmetric pieces normal to two radial directions and with a vertex angle $\pi-\alpha$ between the cutting lines, as illustrated in Fig.\ \ref{fig:fusion}. This simple setup enables us to manipulate the fusion of adjacent MZMs in different ways.

We are able to control the fusion by tuning the direction of magnetic field
$\theta$, depending on the value of $\alpha$ which measures the
range of angles missing in the setup. If $\alpha>\text{max}(\varphi_{s1},\varphi_{s2})$, then we can move any two adjacent MZMs into the vertex and fuse them by tuning $\theta$, see Fig.\ \ref{fig:fusion}(a). Take the
reflection-symmetry line of the incomplete disk in the $x$ direction
for instance. At $\theta=0$, $\pi/4$, $\pi/2$ and $3\pi/4$ (modulo
$\pi$), respectively, the adjacent pairs of $\gamma_{1/2}$, $\gamma_{4/1}$,
$\gamma_{3/4}$, and $\gamma_{2/3}$\textbf{ }are maximally fused.
Note that the fusion between $\gamma_{1/4}$ ($\gamma_{2/3}$)\textbf{
}always requires a finite $\mu$. If $\text{min}(\varphi_{s1},\varphi_{s2})<\alpha<\text{max}(\varphi_{s1},\varphi_{s2})$,
then only the adjacent pairs with angular separation given by $\text{min}(\varphi_{1s},\varphi_{2s})$
can be fused. Whereas the other adjacent pairs with angular separation
$\text{max}(\varphi_{s1},\varphi_{s2})$ can never be pushed to the
vertex together and their fusion is suppressed. Finally, if $\alpha<\text{min}(\varphi_{s1},\varphi_{s2})$, then we cannot fuse the MZMs at all.

Since the separations $\varphi_{s1}$ and $\varphi_{s2}$ depend on the ratio $\bar{\Delta}/B$,
we can also modulate the fusion by tuning $B$ or $\mu$. As an illustration,
we first tune ${\bf B}$ with the field direction set at $\theta=0$, such
that two MZMs, say $\gamma_{1}$ and $\gamma_{2}$, are brought close
to the vertex, see Fig.\ \ref{fig:fusion}(b). When $B>B_{c1}\equiv\bar{\Delta}/\sin(\alpha/2)$,
the separation $\varphi_{1s}$ between $\gamma_{1/2}$ is larger than
$\alpha$. Thus, $\gamma_{1}$ and $\gamma_{2}$ remain well separated.
In contrast, when $B<B_{c1}$, we find $\varphi_{1s}<\alpha$. Then, $\gamma_{1}$
and $\gamma_{2}$ collide and fuse at the vertex. At the critical
point $B=B_{c1}$, $\gamma_{1}$ and $\gamma_{2}$ extend along the
cutting lines, see the middle panel of Fig.\ \ref{fig:fusion}(b).
In all the cases, the other MZMs remain well separated and stay at
zero energy. Similarly, we can set ${\bf B}$ to other proper directions,
e.g., $\theta=\pi/4$, $\pi/2$ or $3\pi/4$, and control the fusion of
other adjacent pairs by tuning $B$. For the pair of $\gamma_{1/4}$
(or $\gamma_{2/3}$), the critical field is given by $B_{c2}=\bar{\Delta}/\cos(\alpha/2)$.
The fusion is achieved when $B>B_{c2}$. Alternatively, we can fix
${B}$ and $\theta$ and instead use $\mu$ to electrically control
the fusion. For the pair of $\gamma_{1/2}$ (or $\gamma_{3/4}$),
the fusion is enhanced when $|\mu|>\mu_{c1}$ and suppressed when
$|\mu|<\mu_{c1}$, where $\mu_{c1}=[B^{2}\sin^{2}(\alpha/2)-\Delta_{0}^{2}]^{1/2}$,
see Fig.\ \ref{fig:fusion}(c). For the pair of $\gamma_{1/4}$ (or
$\gamma_{2/3}$), the critical $\mu$ reads $\mu_{c2}=[B^{2}\cos^{2}(\alpha/2)-\Delta_{0}^{2}]^{1/2}$
and the fusion is reduced when $|\mu|>\mu_{c2}$ whereas revived when
$|\mu|<\mu_{c2}$.

\begin{figure}[bp]
\centering

\includegraphics[width=1\columnwidth]{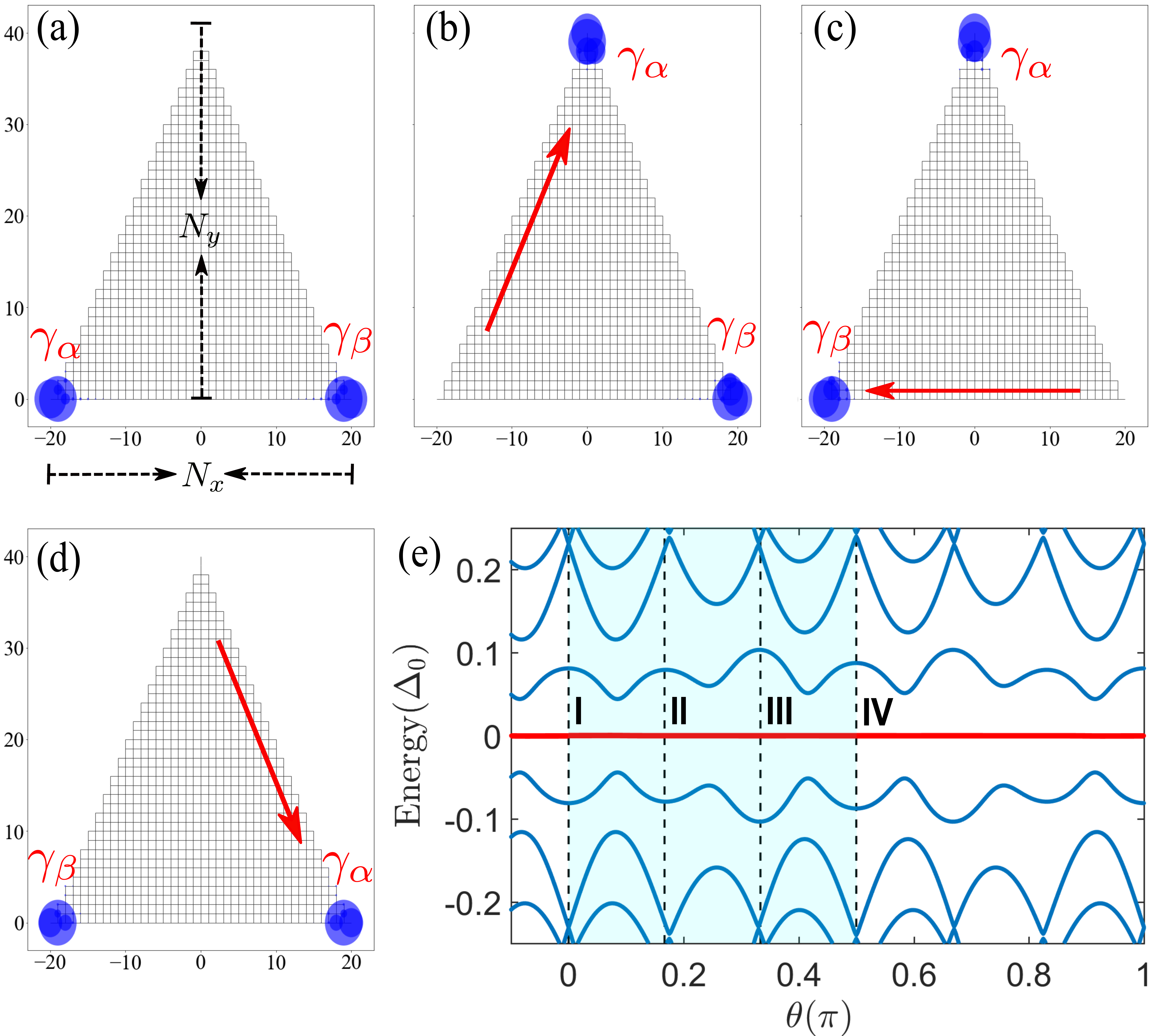}

\caption{Numerical simulation of exchanging two MZMs by varying $\theta$. (a)-(d) Positions of the two MZMs (blue density dots) at $\theta=0$, $\pi/6$, $\pi/3$ and $\pi/2$, respectively. (e) Energy spectrum with respect to $\theta$. There are two zero MZMs, $\gamma_{\alpha}$ and $\gamma_{\beta}$ (red curves). (a-d) correspond to the four subsequent snapshots at $\bf{I}$-$\bf{IV}$, respectively. Increasing $\theta$ by $\pi/2$, $\gamma_{\alpha}$ and $\gamma_{\beta}$ exchange their positions effectively. We provide an animation for this process in the supplemental material\ \citep{SuppInf}. The dimensions of the triangle are $N_{x}=N_{y}=40a$, $\mu=0.1m_{0}$, and all other parameters are the same as those in Fig.\ \ref{fig:MajoranaModes-complete-disk}.}
\label{fig:fusion-triangle}
\end{figure}

\section{Braiding Majorana zero modes and non-Abelian statistics} \label{sec:braiding}
\subsection{Braiding two Majorana zero modes}

We have shown that there are four MZMs on the boundary of a disk. By geometry engineering, i.e., line cutting, and making certain angular positions unavailable, we are able to selectively squeeze two MZMs into the same position and fuse them there. In this section, we show that a single triangle setup with a finite $\mu$ enables us to fuse two out of the four MZMs, while the other two modes remain at zero energy. Such a setup allows us to exchange the remaining two MZMs by rotating the magnetic field ${\bf B}$, without interference from the other modes. The three angles of the triangle are assumed to be smaller than $\text{min}(\varphi_{s1},\varphi_{s2})$. We label the remaining MZMs by $\gamma_{\alpha}$ and $\gamma_{\beta}$, respectively, and discuss the exchange process based on numerical simulations below.

\begin{figure*}[htp]
\includegraphics[width=0.95\textwidth]{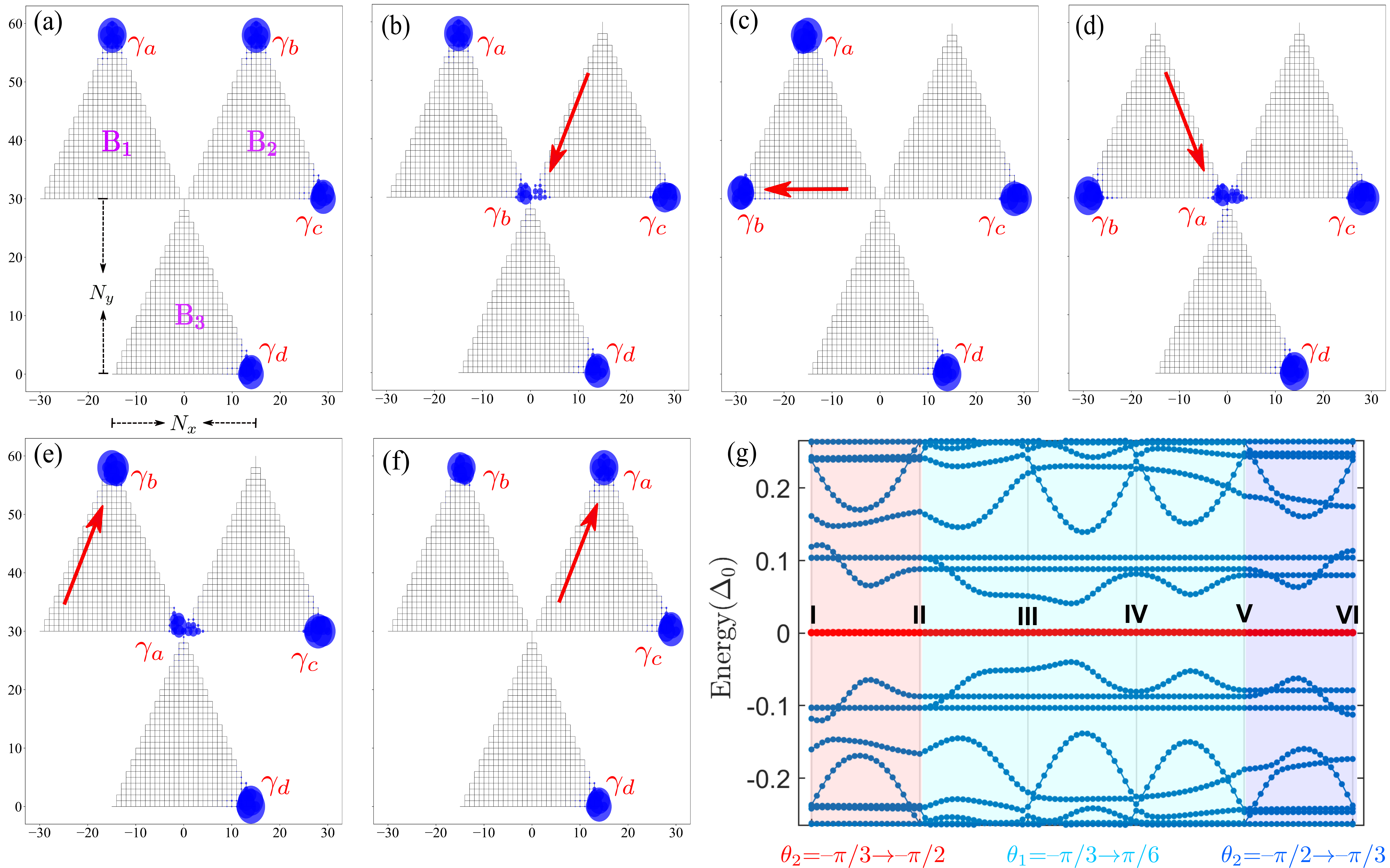}

\caption{Numerical simulation of exchanging $\gamma_a$ and $\gamma_b$ by rotating in-plane magnetic fields. Positions of the four MZMs at (a) ${\bm{\theta}}\equiv(\theta_{1},\theta_{2},\theta_{3})=-(\pi/3,\pi/3,\pi/3)$, (b) ${\bm{\theta}}=-(\pi/3,\pi/2,\pi/3)$, (c) ${\bm{\theta}}=-(\pi/6,\pi/2,\pi/3)$, (d) ${\bm{\theta}}=-(0,\pi/2,\pi/3)$, (e) ${\bm{\theta}}=(\pi/6,-\pi/2,-\pi/3)$, and
(f) ${\bm{\theta}}=(\pi/6,-\pi/3,-\pi/3)$, respectively. (g) Energy spectrum of the process. The three light colored areas indicate the three steps described in the main text. (a-f) correspond to the six subsequent snapshots at $\bf{I}$-$\bf{VI}$, respectively. There are four MZMs (red curves) separated by an energy gap from other modes (blue curves). We provide an animation for the process in the supplemental material\ \citep{SuppInf}. The dimensions of the triangles are $N_{x}=N_{y}=30a$, the pairing phases are $\Phi_{1}=0$, $\Phi_{2}=-\Phi_{3}=2\pi/3$, and all other parameters are the same as those in Fig.\ \ref{fig:fusion-triangle}.}
\label{fig:fusion-three-Triangle}
\end{figure*}

Let us start from the state with the magnetic field applied in the $x$ direction, i.e., $\theta=0$. In the initial state, $\gamma_{\alpha}$ sits at the left (bottom) vertex while $\gamma_{\beta}$ sits at the right (bottom) vertex, see Fig.\ \ref{fig:fusion-triangle}(a). We adiabatically increase $\theta$ by $\pi/2$. As a consequence, the following movements occur in sequence. First, $\gamma_{\beta}$ smoothly shifts to the
top vertex, while $\gamma_{\alpha}$ stays at the left vertex, see Fig.\ \ref{fig:fusion-triangle}(b). Then, $\gamma_{\beta}$ stays at the top vertex while $\gamma_{\alpha}$ shifts to the right vertex, where $\gamma_{\beta}$ was sitting in the initial state. Finally, $\gamma_{\alpha}$ stays at the right vertex, while $\gamma_{\beta}$ shifts down to the left vertex. We can clearly trace the movements of each MZM \cite{SuppInf}. Apparently, $\gamma_{\alpha}$ and $\gamma_{\beta}$ exchange their positions. It is important to note that, during the entire process, $\gamma_{\alpha}$ and $\gamma_{\beta}$ robustly stay at zero energy and they are protected from mixing with other modes by an energy gap, as shown by the instantaneous spectrum in Fig.\ \ref{fig:fusion-triangle}(e). It indicates that the degeneracy of the ground states remains unchanged, which is a necessary condition for an adiabatic operation.

 We now investigate how the Majorana exchange operations affect the quantum states of the triangle setup. It is instructive to inspect the vertices of the triangle on the boundary of a disk and identify $\gamma_\alpha$ and $\gamma_\beta$ with the four MZMs in the disk geometry (see App.\ \ref{sec:Non-Abelian-Manipulations} for more details). By comparing the wavefunctions of $\gamma_\alpha$ and $\gamma_\beta$ in the initial and final states, we find that these two MZMs evolve differently. The two MZMs can be combined to define a complex fermion. Thus, there are two degenerate ground states of different fermion parity, which correspond to the presence and absence of the fermion, respectively. Then, the different evolutions of MZMs indicate that the exchange operation acts diagonally and transforms the two ground states in different ways. After the operation, the Hamiltonian of the system has changed. Consequently, the ground-state manifold is different. This can be attributed to the presence of different MZMs (distinguished by their intrinsic and unique spinor form of the wavefunction) in the SOTS, in contrast to the vortex-bounded MZMs \cite{Ivanov01PRL} which are essentially spinless. Interestingly, when exchanging the MZMs twice, we find that the ground states are transformed into their inversion symmetry counterparts. To revert the system to the original ground-state manifold, we should keep rotating the magnetic field until, after a $2\pi$ rotation, it points again in the initial direction. This corresponds to exchanging the MZMs four times. We can find numerically that after the four-time exchange process, both MZMs accumulate a $\pi$ Berry phase and flip sign,
\begin{eqnarray}
  \gamma_\alpha \mapsto -\gamma_\alpha \ \text{and}\ \gamma_\beta \mapsto -\gamma_\beta.
\end{eqnarray}
This feature can be exploited for topological quantum computation, as discussed below.

\subsection{Braiding more Majorana zero modes\label{sec:braiding-more-modes}}
To exploit the non-Abelian statistics of braiding operations for topological quantum computation, more than two MZMs are required \citep{Bravyi06PRA}. To this end, we consider a trijunction made by connecting three triangle islands with different pairing phases in the center, as illustrated in Fig.\ \ref{fig:fusion-three-Triangle}. The triangle islands would each host a couple of MZMs at the vertices if they were not connected, as we have shown above. However, as the islands have their vertices connected in the center, two MZMs are fused. Therefore, totally only four MZMs remain in the setup. When the magnetic fields are applied uniformly, two of them are in one island and the other two in the other two islands,
respectively. We denote the islands as $T1$, $T2$ and $T3,$ and the MZMs as $\gamma_{a}$, $\gamma_{b}$, $\gamma_{c}$ and $\gamma_{d}$,
respectively. Assume that three magnetic fields can be independently tuned in the three islands. With these preconditions, we show that
this trijunction allows to braid any two of the four MZMs while keeping the other MZMs untouched. The two MZMs in the same island can be braided
by rotating the corresponding magnetic field in the same way as described in the previous section. Thus, we focus on the braiding of two MZMs
from different islands in the following.

Let us start by considering the situation where the magnetic fields ${\bf B}_{i}$ are applied uniformly, i.e., $B_{i}=B_{>}(>\bar{\Delta})$ and $\theta_{i}=-\pi/3$ (with $i\in\{1,2,3\}$). Under this condition, $\gamma_{a}$ is located in $T1$, $\gamma_{b}$ and $\gamma_{c}$ in $T2$, and $\gamma_{d}$ in $T3$, as shown in Fig.\ \ref{fig:fusion-three-Triangle}(a). We braid $\gamma_{a}$ and $\gamma_{b}$ in three steps in sequence.
In the first step, we turn $\theta_{2}=-\pi/3\rightarrow-\pi/2$ and move $\gamma_{b}$ slowly to the center. After this step, $T1$ has
two MZMs, $\gamma_{a}$ and $\gamma_{b}$, see Fig.\ \ref{fig:fusion-three-Triangle}(b). In the second step, we increase $\theta_{1}$ by $\pi/2$. This results in the exchange of $\gamma_{a}$ and $\gamma_{b}$ inside $T1$, see Fig.\ \ref{fig:fusion-three-Triangle}(b)-(e). In the last step, we rotate $\theta_{2}$ back to $-\pi/3$. Thus, $\gamma_{a}$ is moved smoothly to the top vertex of $T2$. Therefore, the positions of $\gamma_{a}$ and $\gamma_{b}$ are exchanged. Importantly, during the entire process the four MZMs stay robustly at zero energy and they are also protected from high-energy modes by a finite energy gap, see Fig.\ \ref{fig:fusion-three-Triangle}(g). By repeating the same procedure, we can exchange $\gamma_a$ and $\gamma_b$ twice, three times, or more times. In a similar way, we can also braid $\gamma_{c}$ and $\gamma_{d}$ by rotating $\theta_{2}$ and $\theta_{3}$ alternatively (see App.\ \ref{sec:basic-braiding} for details). The three exchanges $\gamma_{a}\leftrightarrow\gamma_{b}$, $\gamma_{b}\leftrightarrow\gamma_{c}$ and $\gamma_{c}\leftrightarrow\gamma_{d}$ generate the whole braid group of the four MZMs.

Next, we discuss the interpretation of the braiding in terms of quantum gates. There are totally four MZMs in the setup. We may combine $\gamma_{b}$ and $\gamma_{c}$ to define a complex fermion as $f_{bc}=(\gamma_{b}+i\gamma_{c})/2$. Analogously, we define another complex fermion by combining the other two MZMs, $\gamma_a$ and $\gamma_d$, as $f_{ad}=(\gamma_{d}+i\gamma_{a})/2$. For a fixed global fermion parity, there are two ground states which span the computational space of a single non-local qubit. Without loss of generality, we write the two ground states as $|00\rangle$ and $|11\rangle$, where $n_{bc/ad}\in\{0,1\}$ in the Dirac notation $|n_{bc}n_{ad}\rangle$ are the occupation numbers of the fermions $f_{bc}$ and $f_{ad}$, respectively. For quantum computation, it is essential for the system to revert to the initial configuration each time after a completed braiding so that the ground-state manifold remain the same. Therefore, we have to consider the operations of exchanging MZMs four times in succession. This flips both signs of the two exchanged MZMs as we discussed in the previous section. We find that the four-time exchange of $\gamma_{a}$ and $\gamma_{b}$ corresponds to a $\sigma_x$ gate on the qubit (see App.\ \ref{sec:Non-Abelian-Manipulations} for more details). The same result is obtained by exchanging $\gamma_{c}$ and $\gamma_{d}$ four times. Similarly, the exchange $\gamma_{b}\leftrightarrow\gamma_{c}$ (or $\gamma_{a}\leftrightarrow\gamma_{d}$) corresponds to a $\sigma_z$ gate. These operations are clearly non-commutative, consistent with the non-Abelian nature of the MZMs.

\begin{figure}[b]
\includegraphics[width=1\columnwidth]{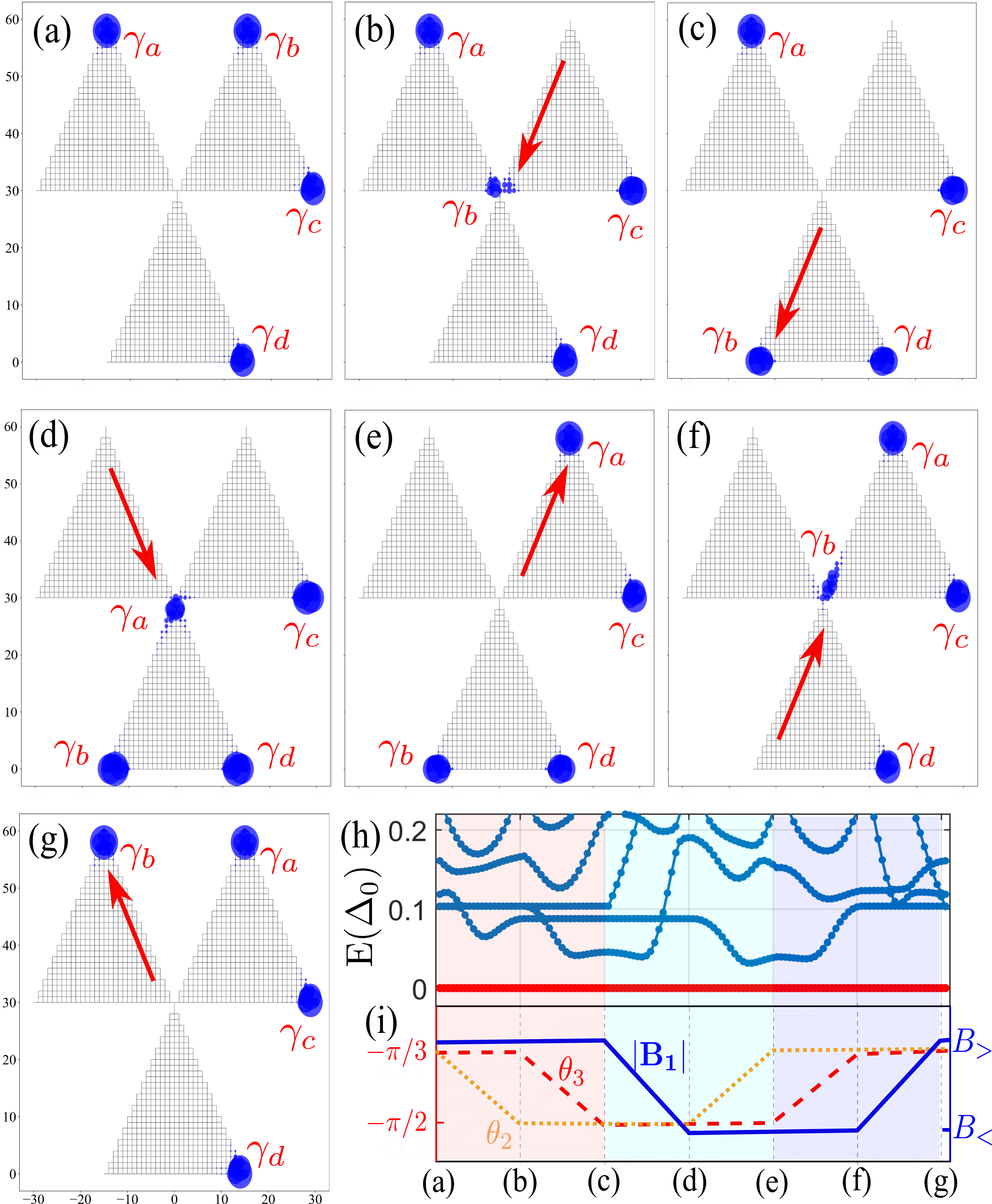}

\caption{Numerical simulation of braiding $\gamma_{a}$ and $\gamma_{b}$.  (a-g) Positions of the MZMs at several subsequent instants. During the protocol, $\theta_{2}$, $\theta_{3}$ and $B_1$ are varied in time, according to (i), while $\theta_1=-\pi/3$ and all other parameters are fixed as the same as those in Fig.\ \ref{fig:fusion-three-Triangle}. (h) The energy spectrum of the system during the process. It is symmetric with respect to zero energy. For the numerical simulations, we use $B_<=0.3m_0$ and $B_>=0.6m_0$.}

\label{fig:T-junction}
\end{figure}

So far, we have only discussed the braiding of MZMs by purely rotating the magnetic fields. In this way, in order to flip both signs of the two exchanged MZMs, we need to exchange them four times. In fact, we are able to flip both signs of two MZMs by only exchanging them twice if we not only rotate the directions but also vary the strengths of the magnetic fields. For illustration, let us consider the same initial configuration as in Fig.\ \ref{fig:fusion-three-Triangle} and take the exchange $\gamma_a\leftrightarrow\gamma_b$ as an example. We can also summarize the operations in three steps. In the first step, we rotate the magnetic field $\bf{B}_2$ from $\theta_2=-\pi/3$ to $-\pi/2$ and then rotate $\bf{B}_3$ from $\theta_3=-\pi/3$ to $-\pi/6$. This moves $\gamma_b$ to the left-bottom vortex via the center, see Fig.\ \ref{fig:T-junction}(b) and (c). In the second step, we move $\gamma_a$ to the right-top vortex. To do so, we decrease the strength of $\bf{B}_1$ to a value $B_<$ smaller than $\bar{\Delta}\equiv\sqrt{\Delta^2_0+\mu^2}$, and subsequently rotate $\bm{B}_2$ back to $\theta=-\pi/3$, see Fig.\ \ref{fig:T-junction}(d) and (e). Finally, we rotate $\bf{B}_3$ back to $\theta_3=-\pi/3$ and increase the strength $\bf{B}_1$ back to $B_> (>\bar{\Delta})$, see Fig.\ \ref{fig:T-junction}(f) and (g). Eventually, $\gamma_a$  and $\gamma_b$ exchange their positions.

In this protocol, our trijunction setup, in fact, mimics a $T$-junction\cite{Alicea11NP}. Then, our MZMs behave similarly to vortices in $p$-wave superconductors \cite{Read00PRB,Ivanov01PRL}. Importantly, the system reverts to the initial configuration after every exchange. The single exchange operation on two MZMs adds a $\pi$ Berry-phase difference to the MZMs. Define the same qubit as before: $\{|00\rangle,|11\rangle\}$, where $n_{bc/ad}\in\{0,1\}$ in the notation $|n_{bc}n_{ad}\rangle$ denote the occupation numbers of the fermions $f_{bc}=(\gamma_b+i\gamma_c)/2$ and $f_{ad}=(\gamma_d+i\gamma_a)/2$, respectively. We can find that the exchange $\gamma_a\leftrightarrow\gamma_b$ or $\gamma_c\leftrightarrow\gamma_d$ corresponds to a $\pi/2$ rotation in the Bloch sphere around the $x$ axis and mixes the qubit. In contrast, the exchange $\gamma_b\leftrightarrow\gamma_c$ or $\gamma_a\leftrightarrow\gamma_d$ corresponds to a $\pi/2$ rotation around the $z$ axis. It acts diagonally on the qubit.

The triangle setups and schemes described above can be further generalized to the case with more pairs of MZMs. As a natural generalization, we consider a ladder structure formed by $N_{\text{tri}}$ elementary triangle islands $T_{i}$ with finite chemical potential, as sketched in Fig.\ \ref{fig:main-results}(c). $N_{\text{tri}}$ is assumed to be an odd integer. Three islands connected by the same point have different pairing phases. For example, we may use $\Phi_{3n-2}=0$ and $\Phi_{3n-1}=-\Phi_{3n}=2\pi/3$ with $n\in\{1,2,...\}$. For uniform magnetic fields, say, with $B_i=B_>$ and $\theta_{i}=-\pi/3$, this ladder setup supports $N_{\text{tri}}+1$ MZMs in total. The setup can be scaled up by adding extra islands. We can braid any two of the MZMs by alternately varying the magnetic fields in analogous ways as described before in the triangle setups. Notably, since the braiding operations are topological, the choice of chemical potential and pairing phases and the controlling of magnetic fields do not need to be fine tuned.

\section{Holonimic gates} \label{sec:holonomic}
According to the Gottesman-Knill theorem
\citep{Gottesman99book}, the topological quantum gates provided by
Majorana braiding are not sufficient for universal quantum computation.
The latter requires indeed at least one non-Clifford single-qubit
gate, such as the T gate (also known as the magic gate) \citep{Bravyi05PRA,Nielsen02Book},
which cannot be implemented just by exchanging MZMs \citep{Bravyi06PRA}.
In general, non-topological procedures for realizing a Majorana magic
gate require a very precise control of the system parameters and,
as a result, they heavily rely on conventional error-correction schemes
\citep{Bravyi05PRA}. In this respect, an interesting role is played
by holonomic gates \citep{Zanardi99PPA} which, while
not being topologically protected, can still feature a good degree
of robustness against errors, thus reducing the amount of hardware
necessary for subsequent state distillation \citep{Karzig16PRX}.

The minimal model for Majorana-based holonomic gates has been proposed
by Karzig \textit{et al} \cite{Karzig16PRX} and consists of four "active"
MZMs, $\gamma_{x}$, $\gamma_{y}$, $\gamma_{z}$ and $\gamma_{0}$. To successfully implement a holonomic gate, one has to adiabatically
vary the three coupling strengths $t_{i}$ between $\gamma_{i}$ (with
$i\in\{x,y,z\}$) and $\gamma_{0}$. Indeed, for every closed loop
in the three-dimensional parameter space spanned by $(t_{x},t_{y},t_{z})$, a difference between the Berry phases picked up by the two states
is developed. By properly designing the loop, it is possible to implement
a T gate and take advantage of a universal geometrical decoupling
in order to suppress the effect of finite control accuracy on the
couplings $t_{i}$ \citep{Karzig16PRX}. One of the main sources of
errors is represented by the parasitic couplings $t_{ij}$ between
$\gamma_{i}$ and $\gamma_{j}$ (with $i,j\in\{x,y,z\}$ and $i\neq j$).
These couplings, which are basically unavoidable in a quantum wire-based
setup, introduce additional and unwanted dynamical phases, which reduce
the gate fidelity (despite error mitigation techniques based on conventional
echo schemes) \citep{Karzig16PRX,Karzig19PRB}.

\begin{figure}[htp]
\centering
\includegraphics[width=1\columnwidth]{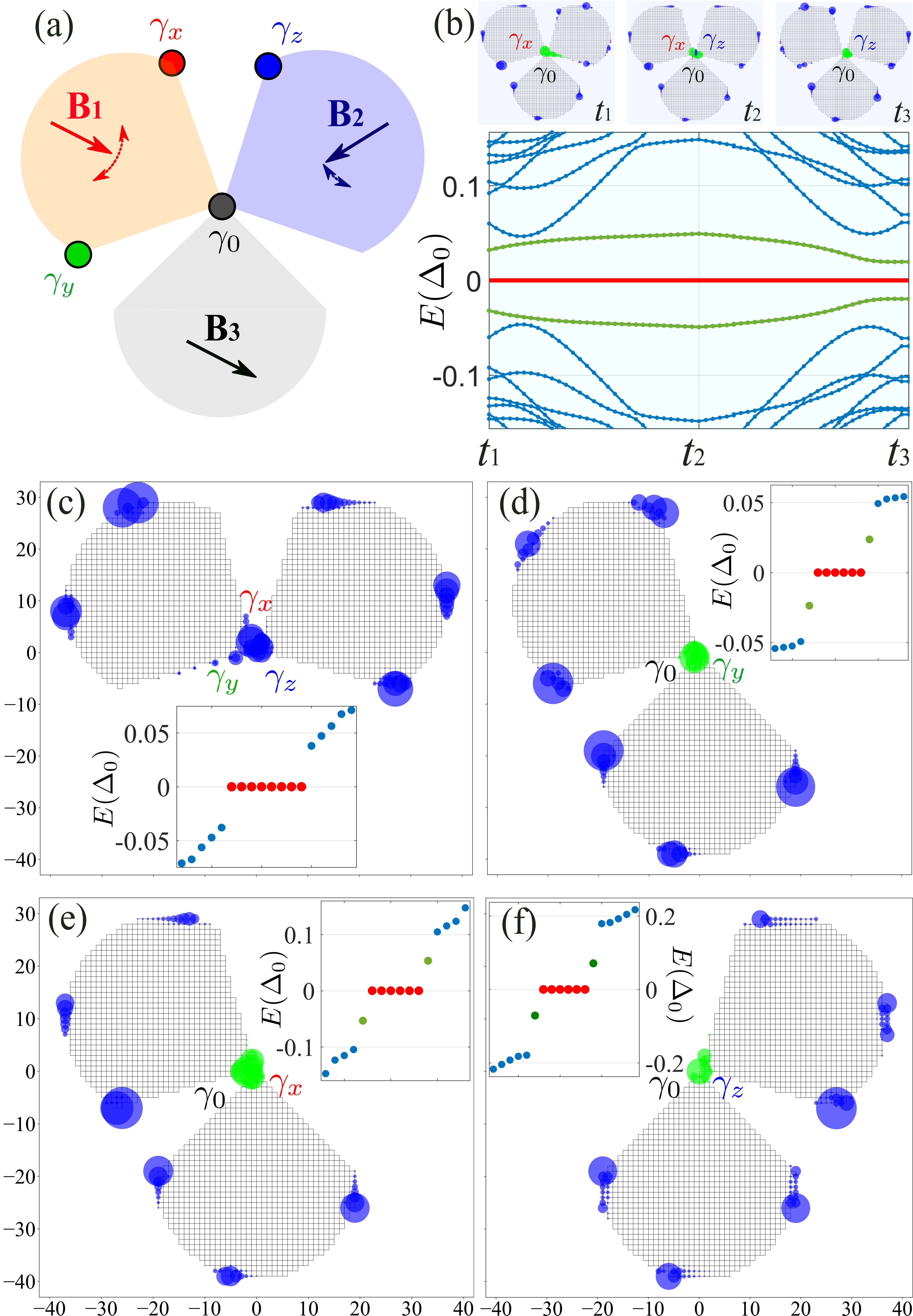}
\caption{Holonomic gates based on the MZMs. (a) Schematics of the four MZMs $\gamma_{x}$, $\gamma_{y}$, $\gamma_{z}$ and $\gamma_{0}$
relevant for the holonomic gates in a shamrock-like trijunction. The positions of these modes are controllable by the magnetic fields ${\bf B}_{1}$, ${\bf B}_{2}$ and ${\bf B}_{3}$. (b) Energy spectrum of the process that first moves $\gamma_{z}$ to the center (during $t_{1}\rightarrow t_{2}$) and then moves $\gamma_{x}$ away from the center (during $t_{2}\rightarrow t_{3}$). Insets are the positions of MZMs in the initial ($\theta_{1}=-\pi/3$ and $\theta_{2}=7\pi/6$), middle ($\theta_{1}=-\pi/3$ and $\theta_{2}=4\pi/3$) and final ($\theta_{1}=-\pi/6$ and $\theta_{2}=4\pi/3$) states. The green densities are for the fused modes (i.e., two excited modes with lowest energy). (c-f) Coupling between different MZMs, i.e., $\gamma_{x}$, $\gamma_{y}$ and $\gamma_{z}$ in (c); $\gamma_{y}$ and $\gamma_{0}$ in (d); $\gamma_{x}$ and $\gamma_{0}$ in (e); and $\gamma_{z}$ and $\gamma_{z}$ in (f). The corresponding energy spectra are displayed in the insets.  $B_{1}=0.54m_{0}$ in (a,c-f) and $0.65m_{0}$ in (a). In (c) $\theta_{1}=-\pi/6$, $\theta_{2}=4\pi/3$, (d) $\theta_{1}=0$, (e) $\theta_{1}=-\pi/3$, (f) $\theta_{2}=4\pi/3$, for all panels, $\Phi_{1}=\Phi_{2}=0$, $\Phi_{3}=2\pi/3$, $B_{2}=B_{3}=0.54m_{0}$, $\theta_{3}=-\pi/3$, $\mu=0$, $\Delta_{0}=0.4m_{0}$, $m_{0}=2$, $N_{r}=20a$, and all other parameters are the same as those in Fig.\ \ref{fig:MajoranaModes-complete-disk}.}
 \label{fig:holonomic-gates}
\end{figure}

Remarkably, SOTS-based setups can naturally guarantee vanishing parasitic
couplings, thus providing a novel and convenient platform to study
and implement Majorana-based holonomic gates. By exploiting the effective
mirror symmetry featured by the SOTSs at $\mu=0$, one can indeed
have $t_{ij}=0$ while still being able to smoothly and freely
tune all the other three couplings $t_{i}$. In the following, we
demonstrate this remarkable feature with a concrete example.
The setup we propose is a shamrock-like trijunction which is formed
by connecting the vertices of three incomplete disks (denoted as $S1$,
$S2$ and $S3$, respectively) with vanishing $\mu$, as sketched
by Fig.\ \ref{fig:holonomic-gates}(a). When the three incomplete disks
are disconnected, they each host four MZMs labeled by $\gamma_{i}$
with $i\in\{1,2,3,4\}$. The positions of the MZMs are controllable
by the applied magnetic fields, as discussed before. For concreteness,
we consider the incomplete disks with the same vertex angle of $\pi/2$
and evenly distributed. By properly tuning the magnetic fields ${\bf B}_{1}$ and ${\bf B}_{2}$, i.e., $B_{1},B_{2}\simeq\sqrt{2}\Delta_{0}$,
$\theta_{1}\simeq-\pi/6$ and $\theta_{2}\simeq7\pi/6$, we move the
MZMs $\gamma_{1}$ and $\gamma_{4}$ of $S1$ and $\gamma_{1}$ of
$S2$ close to or in the center. Due to the effective mirror symmetry,
these three adjacent MZMs cannot interact with each other even if
they have strong overlap in their wavefunctions. We thus identify
them with $\gamma_{x}$, $\gamma_{y}$ and $\gamma_{z}$, respectively.
Whether $S1$ and $S2$ have the same pairing phase or not is not
important. We set both at zero $\Phi_{1}=\Phi_{2}=0$ for concreteness.
On the other hand, we adopt a different pairing phase $\Phi_{3}\neq\{0,\pi\}$
in $S3$, and fix the MZM $\gamma_{2}$ or $\gamma_{3}$ of $S3$
in the center by tuning $\text{{\bf B}}_{3}$. This MZM would fuse
with $\gamma_{i}$ (with $i\in\{x,y,z\}$) if they are brought close
to the center. We denote it as $\gamma_{0}$. The couplings $t_{i}$
between $\gamma_{i}$ and $\gamma_{0}$ are closely determined by their distance. Recalling that these distances are controllable by ${\bf B}_{1}$ or ${\bf B}_{2}$, the couplings are thus smoothly adjustable.

To test our analysis and demonstrate the basic conditions for holonomic gates, we first consider $S1$ and $S2$ together and move $\gamma_{x}$,
$\gamma_{y}$ and $\gamma_{z}$ all to the center, see Fig.\ \ref{fig:holonomic-gates}(c). From the energy spectrum [inset of Fig.\ \ref{fig:holonomic-gates}(c)], we find that there is no splitting of MZMs from zero energy. This clearly indicates that $\gamma_{x}$, $\gamma_{y}$ and $\gamma_{z}$ can never interact with each other. We next consider $S_{3}$ together with $S_{1}$ or $S_{2}$ and move $\gamma_{x},$ $\gamma_{y}$ and $\gamma_{z}$ to the center, respectively, see Fig.\ \ref{fig:holonomic-gates}(d)-(f). In contrast to Fig.\ \ref{fig:holonomic-gates}(c), evident energy splittings of two original MZMs can be observed (green dots in the insets). This signifies the couplings between $\gamma_{0}$ and $\gamma_{x}$, $\gamma_{y}$ and $\gamma_{z}$, respectively. The splitting energies are smooth functions of ${\bf B}_{1}$ or ${\bf B}_{2}$. Thus, by tuning ${\bf B}_{1}$ and ${\bf B}_{2}$, we are able to adiabatically adjust the couplings $t_{i}$. For illustration, we start with the state, where $\gamma_{x}$ and $\gamma_{0}$ are in the center, while the other two not. In this initial state, we have $|t_{x}|\gg|t_{y}|,|t_{z}|$. We slowly move $\gamma_{z}$ also into the center by carefully rotating ${\bf B}_{2}$ and then move $\gamma_{x}$ away from the center by rotating ${\bf B}_{1}$. In the final state, we arrive at $|t_{z}|\gg|t_{x}|,|t_{y}|$. The instantaneous spectrum for this process is displayed in Fig.\ \ref{fig:holonomic-gates}(b). Two finite energy levels resulting from the coupling of $\gamma_{0}$ and $\gamma_{x}$ or $\gamma_{z}$ are always present (green lines), while the other MZMs stay robust at zero energy (red curves). This clearly shows that $t_{x}$ and $t_{z}$ are smoothly varied. Similarly, the variation of ${\bf B}_{1}$ enables us to adjust the other coupling $t_{y}$ in a smooth manner.

Finally, we note that the shamrock-like trijunction can realize the
full parameter space of $(t_{x},t_{y},t_{z})$. In this sense, a full
set of holonomic gates defined by the four coupled MZMs can be achieved.
There are many alternative ways of choosing the four relevant MZMs,
which, however, share the essential physics. Other MZMs that exist
in the setup but are not relevant to the problem can be safely ignored
since they are always far away from the center.

\section{Discussion} \label{sec:discussion}
There are various candidate systems to implement our schemes. Particularly, the inverted HgTe quantum well in proximity to conventional superconductors and monolayer $\text{FeTe}{}_{1-x}\text{Se}_{x}$ have the right properties. The superconducting proximity effect has been realized in HgTe quantum wells \citep{Hart14Nphys,Ren19Nature}. The application of high in-plane magnetic fields is also feasible in experiments \citep{Ren19Nature}. Monolayer $\text{FeTe}{}_{1-x}\text{Se}_{x}$ can host a quantum spin Hall phase coexisting with intrinsic superconductivity \citep{XXWu16PRB,ShiX17SciB,PengXL19PRB}. Importantly, the superconducting gap is comparably large (up to 16.5 meV) \citep{LiFS15PRB}, and it can sustain a large in-plane magnetic
field (up to 45 T) \citep{Salamon16SRep}. The localization length
of the MZMs can be estimated by $\xi=\text{max}(v/\Delta_{0},m/v)$.
For typical parameters $v\simeq1.0$ eV$\cdot\mathring{\text{A}}$,
$\Delta_{0}\simeq1.0$ meV and $m\simeq10$ eV$\cdot\mathring{\text{A}}^{2}$, we derive $\xi\simeq10^{3}$ $\mathring{\text{A}}$. Therefore, the length scales for our setups should be larger than $100$ nm. For a g-factor close to $g\simeq2$, a magnetic field of around $\Delta_{0}/g\mu_{B}\simeq10$
T would be sufficient to induce the MZMs.

The MZMs in our setups can be locally probed, for example, by scanning tunneling microscopy. Their existence could be signified as zero-bias peaks in the tunneling conductance \citep{Flensberg10prb}.
To read out the qubits, we can apply, for instance, the quantum-dot approach proposed by Plugge \emph{et al.}  \citep{plugge2017majorana} We move a pair of MZMs to two neighboring vertices by tuning the magnetic fields and couple them respectively to two quantum dots. One of the two dots is capacitively coupled to a charge sensor. The couplings and the energy levels of the dots are adjustable by gate voltages. Then, the conductance of the charge sensor depends on the fermion parity encoded in the two MZMs. Thus, it can serve to read out the qubit state.
The parity information encoded in the MZMs could be alternatively measured via the Josephson effect. We bring two MZMs from two different islands to a vortex that connects the two islands. Then, the Josephson current flowing between the two islands depends sensitively on the fermion parity and hence can be used to deduce the parity information \cite{Fu08PRL}. As a nontrivial result of the braiding, we would expect to measure different results before and after the braiding.

In summary, we have provided a minimal model for SOTSs and analyzed
the MZMs in a disk geometry. We have systematically investigated
the fusion of the MZMs. Importantly, we have proposed different setups
of SOTSs which allow to implement topological and holonomic
gates via braiding and fusion of MZMs. Our results establish SOTSs
as an ideal platform for scalable and fault-tolerant quantum information technologies.\\

\section{Acknowledgments}
We thank Nicolas Bauer, Hai-Zhou Lu, Benedikt Scharf, and Xianxin Wu for valuable discussion. This work was supported by the DFG (SPP1666 and SFB1170
``ToCoTronics"), the W\"urzburg-Dresden Cluster of
Excellence ct.qmat, EXC2147, project-id 390858490, and the Elitenetzwerk
Bayern Graduate School on ``Topological Insulators".

\appendix
\section{Boundary states in a large disk\label{sec:Edge-states-in}}

To drive the boundary states in a disk geometry, we use polar coordinates: $r=\sqrt{x^{2}+y^{2}},\ \ \varphi=\arctan(y/x)$ and replace
\begin{eqnarray}
\partial_{x}\pm i\partial_{y} & \rightarrow & e^{\pm i\varphi}(\partial_{r}\pm ir^{-1}\partial_{\varphi}),\nonumber \\
\partial_{x}^{2}+\partial_{y}^{2} &  \rightarrow & \partial_{r}^{2}+r^{-2}\partial_{\varphi}^{2}+r^{-1}\partial_{r}
\end{eqnarray}
in the low-energy Hamiltonian. Take the part for spin-up electrons for illustration. The corresponding Dirac Hamiltonian is given by
\[
\begin{aligned}h_{0}(r,\varphi)= & \begin{pmatrix}m(\partial^{2}) & -ve^{-i\varphi}(i\partial_{r}+r^{-1}\partial_{\varphi})\\
-ve^{i\varphi}(i\partial_{r}-r^{-1}\partial_{\varphi}) & -m(\partial^{2})
\end{pmatrix},\end{aligned}
\]
where $m(\partial^{2})=m_{0}+m(\partial_{r}^{2}+r^{-2}\partial_{\varphi}^{2}+r^{-1}\partial_{r})$.
In this disk model, the angular momentum $\nu$ is a good quantum
number. For a large disk, we can assume an ansatz for the boundary-state wavefunction as
\begin{align}
\psi(r,\varphi) & =e^{i\nu\varphi}e^{\lambda r}(\alpha,\beta e^{i\varphi})^T/\sqrt{r},
\label{eq:trial}
\end{align}
where $\text{Re}(\lambda)\gg1/R$ and $R$ is the radius of the disk.
The $\varphi$ periodicity of the wavefunction $\psi(r,\varphi)=\psi(r,\varphi+2\pi)$
imposes the constraint $\nu\in\mathbb{Z}$. Plugging this ansatz into the Dirac equation, we obtain the eigen equation
\begin{align}
\begin{pmatrix}m_{\nu}+m\lambda^{2}-\epsilon_{\nu} & -iv[\lambda+(\nu+1/2)/r]\\
-iv[\lambda-(\nu+1/2)/r] & -m_{\nu}-m\lambda^{2}-\epsilon_{\nu}
\end{pmatrix}\begin{pmatrix}\alpha\\
\beta
\end{pmatrix} & =0,\label{eq:eigen-equation}
\end{align}
where $m_{\nu}=m_{0}-m(\nu+1/2)^{2}/r^{2}$ and $\epsilon_{\nu}=\epsilon-m(\nu+1/2)/r^{2}$.
A nontrivial solution of $(\alpha,\beta)^{T}$ to Eq.\ (\ref{eq:eigen-equation})
yields
\begin{eqnarray}
(m_{\nu}+m\lambda^{2})^{2}-v^{2}[\lambda^{2}-(\nu+1/2)^{2}/r^{2}] & = & \epsilon_{\nu}^{2}.\label{eq:eigen-equation-2}
\end{eqnarray}
Solving this, we find four solutions of $\lambda$ as $\pm\lambda_{1/2}$ with
\begin{eqnarray}
\lambda_{1/2}^{2} &= &\dfrac{(\nu+1/2)^{2}}{r^{2}}-\dfrac{2mm_{0}-v^{2}}{2m^{2}} \nonumber\\
& & \pm\dfrac{\sqrt{v^{4}-4mm_{0}v^{2}-4m^{2}\epsilon_{\nu}^{2}}}{2m^{2}}.\label{eq:lambda}
\end{eqnarray}
Each $\lambda$ corresponds to a spinor solution of $(\alpha,\beta)^{T}$. We are interested in boundary states whose wavefunctions decay exponentially when away from the boundary to the origin, we assume $\text{Re}[\lambda_{1/2}(R)]>0$ and expand the general wavefunction for boundary states as
\begin{eqnarray}
\Psi_{e\uparrow}(r,\varphi) & = & \sum_{\lambda}C_{\lambda}e^{i\nu\varphi}\dfrac{e^{\lambda r}}{\sqrt{r}}\begin{pmatrix}iv[\lambda+(\nu+1/2)/r]\\
\left(m_{\nu}+m\lambda^{2}-\epsilon_{\nu}\right)e^{i\varphi}
\end{pmatrix}.
\end{eqnarray}

We impose the open boundary conditions to the wavefunction,
\begin{equation}
\Psi_{e\uparrow}(r=R,\varphi)=0.\label{eq:Bounday-condition}
\end{equation}
This leads to the secular equation for nontrivial solutions of $\{C_{\lambda_{1}}e^{\lambda_{1}R},$ $C_{\lambda_{2}}e^{\lambda_{2}R}\}$:
\begin{align}
 & [\lambda_{1}+(\nu+1/2)/R][m_{\nu}+m\lambda_{2}^{2}-\epsilon_{\nu}]\nonumber \\
= & [\lambda_{2}+(\nu+1/2)/R][m_{\nu}+m\lambda_{1}^{2}-\epsilon_{\nu}].
\end{align}
This, together with the expressions of $\lambda_{1}$ and $\lambda_{2}$
in Eq.\ (\ref{eq:lambda}), determines the eigenenergy as
\begin{eqnarray}
\epsilon(\nu) & = & -\text{sgn}(m)|v|\nu/R+m\nu/R^{2}.\label{eq:eigen-value}
\end{eqnarray}
Plugging $\epsilon(\nu)$ back into Eq.\ (\ref{eq:Bounday-condition}), we find $C_{\lambda_{1}}/C_{\lambda_{2}}$. Then, the wavefunctions
of the boundary states can be written as
\begin{eqnarray}
\Psi_{e\uparrow}({\bf x}) & = & e^{i\nu\varphi}K(r)( \text{sgn}(mv),-ie^{i\varphi})^T,
\label{eq:WF}
\end{eqnarray}
where $K'(r)=[e^{\lambda_{1}(r-R)}-e^{\lambda_{2}(r-R)}]/\sqrt{r}$,
${\bf x}\equiv(r,\varphi)$, and
\begin{eqnarray}
\lambda_{1,2} & = & \left|\dfrac{v}{2m}\right|\pm\sqrt{\dfrac{v^{2}}{4m^{2}}-\dfrac{m_{0}}{m}+\dfrac{(v+1/2)^{2}}{R^{2}}}.
\end{eqnarray}

For large $R\gg|m/v|$, we approximate a small segment of the
disk boundary as a straight line. We define effective coordinate and
momentum along the boundary as
\begin{eqnarray}
s\equiv R\varphi,\ \ \ p_{\nu} & \equiv & \nu/R.
\end{eqnarray}
Then, the dispersion (\ref{eq:eigen-value}) becomes
\begin{eqnarray}
\epsilon(p_{\nu}) & = & \text{sgn}(m)|v|p_{\nu},\label{eq:eigen-value-1}
\end{eqnarray}
and the wavefunction (\ref{eq:WF})
\begin{equation}
\Psi_{e\uparrow} = e^{ip_{\nu}s}\dfrac{e^{\lambda_{1}(r-R)}-e^{\lambda_{2}(r-R)}}{\sqrt{R}}\begin{pmatrix}\text{sgn}(mv)\\
-ie^{i\varphi}
\end{pmatrix},\label{eq:WF-1}
\end{equation}
where $\lambda_{1,2}=\left|v/2m\right|\pm\sqrt{v^{2}/4m^{2}-m_{0}/m+p_{\nu}^{2}}.$

Following the same approach, the boundary bands for spin-down electrons, spin-up and spin-down holes are found, respectively, as
\begin{eqnarray}
E_{e,\uparrow}(p_{\nu}) & = & -\text{sgn}(m)|v|p_{\nu}-\mu,\nonumber \\
E_{e,\downarrow}(p_{\nu}) & = & \text{sgn}(m)|v|p_{\nu}-\mu,\nonumber \\
E_{h,\uparrow}(p_{\nu}) & = & -\text{sgn}(m)|v|p_{\nu}+\mu,\nonumber \\
E_{h,\downarrow}(p_{\nu}) & = & \text{sgn}(m)|v|p_{\nu}+\mu.
\end{eqnarray}
Here, we have considered the presence of a finite chemical potential
$\mu$ for generality. Correspondingly, the wavefunctions in the full
basis are given by
\begin{align}
\Psi_{e\uparrow} & =e^{ip_{\nu}s}K(r)(\text{sgn}(mv),-ie^{i\varphi},0,0,0,0,0,0)^{T},\nonumber \\
\Psi_{e\downarrow} & =e^{ip_{\nu}s}K(r)(0,0,\text{sgn}(mv),ie^{-i\varphi},0,0,0,0)^{T},\nonumber \\
\Psi_{h\uparrow} & =e^{ip_{\nu}s}K(r)(0,0,0,0,\text{sgn}(mv),ie^{-i\varphi},0,0)^{T},\nonumber \\
\Psi_{h\downarrow} & =e^{ip_{\nu}s}K(r)(0,0,0,0,0,0,\text{sgn}(mv),-ie^{i\varphi})^{T},
\end{align}
where
 \begin{equation}
K(r)=\mathcal{N}[e^{\lambda_{1}(r-R)}-e^{\lambda_{2}(r-R)}] \label{eq:distribution}
\end{equation}
and $\mathcal{N}$ is the normalization factor. These boundary bands are helical with velocity $v$. They are related
by time-reversal and particle-hole symmetries. Assume $mv>0$ without
loss of generality, we arrive at the wavefunctions stated in the main
text.

\section{Wavefunctions and spin polarizations of MZMs\label{sec:Wavefunction-and-polarization}}

In this section, we derive the wavefunctions of the MZMs. Assume the wavefunction of a given MZM $\gamma_{i}$ near its localization center $\varphi_{i}$ in the form
\begin{equation}
\Psi_{0}=e^{\int \xi ds}(c_{1},c_{2},c_{3},c_{4})^{T},
\end{equation}
where $s=R(\varphi-\varphi_i)$ is measured from $\varphi_{i}$. The eigen equation at zero energy can be written as
\begin{eqnarray}
\begin{pmatrix}iv\xi-\mu & i e^{-i\varphi}\widetilde{B} & 0 & -\Delta_{0}\\
-i e^{i\varphi} \widetilde{B} & -iv\xi-\mu & \Delta_{0} & 0\\
0 & \Delta_{0} & iv\xi+\mu & i e^{i\varphi}\widetilde{B}\\
-\Delta_{0} & 0 & -i e^{-i\varphi}\widetilde{B} & -iv\xi+\mu
\end{pmatrix}\begin{pmatrix}c_{1}\\
c_{2}\\
c_{3}\\
c_{4}
\end{pmatrix} & = & 0, \nonumber
\end{eqnarray}
where $\widetilde{B}=B\sin(\varphi-\theta)$. Accordingly, the secular equation for $\Psi_{0}$ is given by
\begin{align}
\text{det}\begin{pmatrix}iv\xi-\mu & i\widetilde{B} & 0 & -\Delta_{0}\\
-i\widetilde{B} & -iv\xi-\mu & \Delta_{0} & 0\\
0 & \Delta_{0} & iv\xi+\mu & i\widetilde{B}\\
-\Delta_{0} & 0 & -i\widetilde{B} & -iv\xi+\mu
\end{pmatrix}  = & 0. \label{eq:B2}
\end{align}
Solving Eq.\ \eqref{eq:B2}, we find four solutions of $\xi$ as $\pm\xi_{1,2}$
with
\begin{eqnarray}
\xi_{1,2} & = & \Delta_{0}/v\pm\sqrt{\widetilde{B}^{2}-\mu^{2}}/v,
\end{eqnarray}
and the corresponding wavefunctions $\Psi_{0}$.

Let us first consider $\gamma_{1}$ centered at
\begin{eqnarray}
\varphi_{1} & = & \arcsin(\bar{\Delta}/B)+\theta,
\end{eqnarray}
where $\bar{\Delta}=\sqrt{\Delta_{0}^{2}+\mu^{2}}$. For $s$ slightly larger than $\varphi_{1}R$, we have $\text{Re}(-\xi_{1})<0$ and $\text{Re}(\xi_{2})<0$, and for $s$ slightly smaller than $\varphi_{1}R$, we have $\text{Re}(\xi_{1})>0$ and $\text{Re}(\xi_{2})>0$. Thus, the wavefunction around $\varphi_{1}$ can be expanded as
\begin{equation}
\Psi_{1}=\begin{cases}
\alpha_{>}e^{-\int\xi_{1}ds}(i,ie^{i(\varphi_{1}+\vartheta)},e^{i(\varphi_{1}+\vartheta)},1)^{T}\\
+\beta_{>}e^{\int\xi_{2}ds}(-i,-ie^{i(\varphi_{1}-\vartheta)},e^{i(\varphi_{1}-\vartheta)},1)^{T}, & \varphi>\varphi_{1}\\
\alpha_{<}e^{\int\xi_{1}ds}(-i,ie^{i(\varphi_{1}-\vartheta)},-e^{i(\varphi_{1}-\vartheta)},1)^{T}\\
+\beta_{<}e^{\int\xi_{2}ds}(-i,-ie^{i(\varphi_{1}-\vartheta)},e^{i(\varphi_{1}-\vartheta)},1)^{T}, & \varphi<\varphi_{1}
\end{cases} \nonumber
\end{equation}
where $e^{i\vartheta}=(\Delta_{0}+i\mu)/\bar{\Delta}$. Considering the continuity of the wavefunction at $s(\equiv\varphi_{1}R)=0$, we find $\alpha_{>}=\alpha_{<}=0$ and $\beta_{>}=\beta_{<}$. Therefore, $\Psi_{1}$ is given by
\begin{eqnarray}
\Psi_{1} & = & e^{\int\xi_{2}ds}(-i,-ie^{i(\varphi_{1}-\vartheta)},e^{i(\varphi_{1}-\vartheta)},1)^{T}. \label{psi1}
\end{eqnarray}
In the basis of the bulk Hamiltonian \eqref{eq:minimal-model}, the wavefunction reads
\begin{eqnarray}
\Psi_{1} & = & \mathcal{F}_1(e^{-i(\varphi_{1}-\vartheta+\pi/2)/2},-e^{i(\varphi_{1}+\vartheta+\pi/2)/2},\nonumber \\
 &  & \quad e^{i(\varphi_{1}-\vartheta-\pi/2)/2},e^{-i(\varphi_{1}+\vartheta-\pi/2)/2},\nonumber \\
 &  & \quad e^{i(\varphi_{1}-\vartheta+\pi/2)/2},-e^{-i(\varphi_{1}+\vartheta+\pi/2)/2},\nonumber \\
 &  & \quad e^{-i(\varphi_{1}-\vartheta-\pi/2)/2},e^{i(\varphi_{1}+\vartheta-\pi/2)/2})^{T},
\end{eqnarray}
where $\mathcal{F}_1 = e^{i(\varphi_{1}-\vartheta-\pi/2)/2}e^{\int\xi_{2}(s)ds}$ with $K(r)K(r)$ given by Eq.\ \eqref{eq:distribution}.

Next, we consider $\gamma_{2}$ centered at
\begin{eqnarray}
\varphi_{2} & = & -\text{arcsin}(\bar{\Delta}/B)+\theta.
\end{eqnarray}
We expand the wavefunction of $\gamma_{2}$ near $\varphi_{2}$ as
\begin{equation}
\Psi_{2}=\begin{cases}
\alpha_{>}e^{-\int\xi_{1}ds}(i,ie^{i(\varphi_{2}+\vartheta)},e^{i(\varphi_{2}+\vartheta)},1)^{T}\\
+\beta_{>}e^{-\int\xi_{2}ds}(i,-ie^{i(\varphi_{2}+\vartheta)},-e^{i(\varphi_{2}+\vartheta)},1)^{T}, & \varphi>\varphi_{2}\\
\alpha_{<}e^{\int\xi_{1}ds}(-i,ie^{i(\varphi_{2}-\vartheta)},-e^{i(\varphi_{2}-\vartheta)},1)^{T}\\
+\beta_{<}e^{-\int\xi_{2}s}(i,-ie^{i(\varphi_{2}+\vartheta)},-e^{i(\varphi_{2}+\vartheta)},1)^{T}, & \varphi<\varphi_{2}
\end{cases} \nonumber
\end{equation}
Matching $\Psi_{2}(s)$ at $s(\equiv\varphi_{2}R)=0$, $\Psi_{2}(s)$ is found as
\begin{equation}
\Psi_{2}=e^{-\int \xi_{2}s}(i,-ie^{i(\varphi_{2}+\vartheta)},-e^{i(\varphi_{2}+\vartheta)},1)^{T}.
\end{equation}
In the basis of the bulk Hamiltonian, the wavefunction reads
\begin{eqnarray}
\Psi_{2} & = & \mathcal{F}_2(-e^{-i(\varphi_{2}+\vartheta+\pi/2)/2},e^{i(\varphi_{2}-\vartheta+\pi/2)/2},\nonumber \\
 &  & \quad e^{i(\varphi_{2}+\vartheta-\pi/2)/2},e^{-i(\varphi_{2}-\vartheta-\pi/2)/2},\nonumber \\
 &  & \quad-e^{i(\varphi_{2}+\vartheta+\pi/2)/2},e^{-i(\varphi_{2}-\vartheta+\pi/2)/2},\nonumber \\
 &  & \quad e^{-i(\varphi_{2}+\vartheta-\pi/2)/2},e^{i(\varphi_{2}-\vartheta-\pi/2)/2})^{T},
\end{eqnarray}
where $\mathcal{F}_2 = e^{i(\varphi_{2}+\vartheta-\pi/2)/2}e^{-\int\xi_{2}(s)ds}K(r)$.

We now turn to $\gamma_{3}$ which centers at
\begin{eqnarray}
\varphi_{3} & = & \text{arc\ensuremath{\sin}}(\bar{\Delta}/B)+\pi+\theta.
\end{eqnarray}
The wavefunction near $\varphi_{3}$ can be expanded as
\[
\Psi_{3}=\begin{cases}
\alpha_{>}e^{-\int\xi_{1}s}(i,-ie^{i(\varphi_{3}-\vartheta)},-e^{i(\varphi_{3}-\vartheta)},1)^{T}\\
+\beta_{>}e^{\int\xi_{2}s}(-i,ie^{i(\varphi_{3}-\vartheta)},-e^{i(\varphi_{3}-\vartheta)},1)^{T}, & \varphi>\varphi_{3}\\
\alpha_{<}e^{\int\xi_{1}s}(-i,-ie^{i(\varphi_{3}+\vartheta)},e^{i(\varphi_{3}+\vartheta)},1)^{T}\\
+\beta_{<}e^{\int\xi_{2}s}(-i,ie^{i(\varphi_3-\vartheta)},-e^{i(\varphi_{3}-\vartheta)},1)^{T}, & \varphi<\varphi_{3}
\end{cases}
\]
Matching $\Psi_{3}(s)$ at $s(\equiv\varphi_{3}R)=0$, we find
\begin{equation}
\Psi_{3}=e^{\int\xi_{2}ds}(-i,ie^{i(\varphi_{3}-\vartheta)},-e^{i(\varphi_{3}-\vartheta)},1)^{T}. \label{psi3}
\end{equation}
In the basis of the bulk Hamiltonian, the wavefunction reads
\begin{eqnarray}
\Psi_{3} & = & \mathcal{F}_3(e^{-i(\varphi_{3}-\vartheta-\pi/2)/2},e^{i(\varphi_{3}+\vartheta-\pi/2)/2},\nonumber \\
 &  & \quad-e^{i(\varphi_{3}-\vartheta+\pi/2)/2},e^{-i(\varphi_{3}+\vartheta+\pi/2)/2},\nonumber \\
 &  & \quad e^{i(\varphi_{3}-\vartheta-\pi/2)/2},e^{-i(\varphi_{3}+\vartheta-\pi/2)/2},\nonumber \\
 &  & \quad-e^{-i(\varphi_{3}-\vartheta+\pi/2)/2},e^{i(\varphi_{3}+\vartheta+\pi/2)/2})^{T},
\end{eqnarray}
where $\mathcal{F}_3 = e^{i(\varphi_{3}-\vartheta+\pi/2)/2}e^{\int\xi_{2}(s)ds}K(r)$.

Finally, we consider $\gamma_{4}$ at
\begin{eqnarray}
\varphi_{4} & = & -\text{arc\ensuremath{\sin}}(\bar{\Delta}/B)+\pi+\theta.
\end{eqnarray}
The wavefunction near $\varphi_{4}$ is expanded as
\[
\Psi_{4}=\begin{cases}
\alpha_{>}e^{-\int\xi_{1}ds}(i,-ie^{i(\varphi_{4}-\vartheta)},-e^{i(\varphi_{4}-\vartheta)},1)^{T}\\
+\beta_{>}e^{-\int\xi_{2}ds}(i,ie^{i(\varphi_{4}+\vartheta)},e^{i(\varphi_{4}+\vartheta)},1)^{T}, & \varphi>\varphi_{4}\\
\alpha_{<}e^{\int\xi_{1}ds}(-i,-ie^{i(\varphi_{4}+\vartheta)},e^{i(\varphi_{4}+\vartheta)},1)^{T}\\
+\beta_{<}e^{-\int\xi_{2}ds}(i,ie^{i(\varphi_{4}+\vartheta)},e^{i(\varphi_{4}+\vartheta)},1)^{T}, & \varphi<\varphi_{4}
\end{cases}
\]
Matching $\Psi_{4}(s)$ at $s(\equiv\varphi_{4}R)=0$, $\Psi_{4}$
is found as
\begin{equation}
\Psi_{4}= e^{-\int\xi_{2}ds} (i,ie^{i(\varphi_{4}+\vartheta)},e^{i(\varphi_{4}+\vartheta)},1)^{T}.
\end{equation}
In the basis of the bulk model, the wavefunction reads
\begin{eqnarray}
\Psi_{4} & = & \mathcal{F}_4(e^{-i(\varphi_{4}+\vartheta-\pi/2)/2},e^{i(\varphi_{4}-\vartheta-\pi/2)/2},\nonumber \\
 &  & \quad e^{i(\varphi_{4}+\vartheta+\pi/2)/2},-e^{-i(\varphi_{4}-\vartheta+\pi/2)/2},\nonumber \\
 &  & \quad e^{i(\varphi_{4}+\vartheta-\pi/2)/2},e^{-i(\varphi_{4}-\vartheta-\pi/2)/2},\nonumber \\
 &  & \quad e^{-i(\varphi_{4}+\vartheta+\pi/2)/2},-e^{i(\varphi_{4}-\vartheta+\pi/2)/2})^{T},
\end{eqnarray}
where $\mathcal{F}_4 = e^{i(\varphi_{4}+\vartheta+\pi/2)/2}e^{-\int\xi_{2}(s)ds}K(r)$.

The spin operators for electrons and holes are given by
\begin{eqnarray}
\hat{S}_{x}^{(e/h)} & = & (\hbar/4)(\pm\tau_{0}+\tau_{z})s_{x}\sigma_{0},\nonumber \\
\hat{S}_{x}^{(e/h)} & = & (\hbar/4)(\tau_{0}\pm\tau_{z})s_{y}\sigma_{z},\nonumber \\
\hat{S}_{z}^{(e/h)} & = & (\hbar/4)(\pm\tau_{0}+\tau_{z})s_{z}\sigma_{0}.
\end{eqnarray}
The spin polarizations of $\gamma_{1}$ are obtained as
\begin{eqnarray}
S_{\gamma_{1},x}^{(e/h)} & = & \langle\Psi_{1}|\hat{S}_{x}^{(e/h)}|\Psi_{1}\rangle=\pm (\hbar/4)\sin\vartheta\sin\varphi_{1},\nonumber \\
S_{\gamma_{1},y}^{(e/h)} & = & \langle\Psi_{1}|\hat{S}_{y}^{(e/h)}|\Psi_{1}\rangle=\mp (\hbar/4)\sin\vartheta\cos\varphi_{1},\nonumber \\
S_{\gamma_{1},z}^{(e/h)} & = & \langle\Psi_{1}|\hat{S}_{z}^{(e/h)}|\Psi_{1}\rangle=0.
\end{eqnarray}
Similarly, we find the polarizations ${\bf {S}}_{\gamma_i}^{(e/h)} \equiv (S_{\gamma_{i},x}^{(e/h)},$ $S_{\gamma_{i},y}^{(e/h)}, S_{\gamma_{i},z}^{(e/h)})$ for the other three MZMs as
\begin{eqnarray}
{\bf {S}}_{\gamma_2}^{(e/h)} & = & \pm (\hbar\sin\vartheta/4) (\sin\varphi_{2}, - \cos\varphi_{2},0),\nonumber \\
{\bf {S}}_{\gamma_3}^{(e/h)} & = & \pm (\hbar\sin\vartheta/4) (-\sin\varphi_{3}, \cos\varphi_{3},0),\nonumber \\
{\bf {S}}_{\gamma_4}^{(e/h)} & = & \pm (\hbar\sin\vartheta/4) (-\sin\varphi_{4}, \cos\varphi_{4},0).
\end{eqnarray}
The spin polarizations are in the $x$-$y$ plane. They always point parallel or anti-parallel
to the boundary orientation.

\begin{table*}[t]
\caption{Fusion between MZMs on a single or multiple SOTS islands. Here, $\mathcal{F}=v\sin[(\varphi_{I}+\varphi_{II})/2]/2-m\sin[(\varphi_{I}-\varphi_{II})/2]\sin$ and $\mathcal{K}=v\cos[(\varphi_{I}+\varphi_{II})/2]+m\sin[(\varphi_{I}-\varphi_{II})/2]$. The fusion strengths are written in a spinor form: $(|\langle\Psi_{i}|\hat{T}_{x}|\Psi_{j}'\rangle|,|\langle\Psi_{i}|\hat{T}_{y}|\Psi_{j}'\rangle|)$.}

\begin{tabular}{|r@{\extracolsep{0pt}.}l||c|c|c|c|}
\hline
\multicolumn{2}{|c||}{\begin{turn}{90}
\end{turn}} & $\gamma_{1}$ & $\gamma_{2}$ & $\gamma_{3}$ & $\gamma_{4}$\ \ \tabularnewline
\hline
\hline
\multicolumn{2}{|c||}{\ \ $\gamma_{1}'$\ \ } & \textcolor{white}{$\dfrac{}{}$}$(0,0)$ \textcolor{white}{$\dfrac{}{}$} & $(-\mathcal{F},\mathcal{K})\cos\Phi\cos\vartheta$ & $(\mathcal{F}, -\mathcal{K})\sin\Phi$ & $(\mathcal{F}, -\mathcal{K})\sin\vartheta\cos\Phi$ \tabularnewline
\hline
\multicolumn{2}{|c||}{$\gamma_{2}'$} & $\cos\vartheta\cos\Phi(-\mathcal{F},\mathcal{K})$ & \textcolor{white}{$\dfrac{}{}$}$(0,0)$ \textcolor{white}{$\dfrac{}{}$} & $(\mathcal{F},-\mathcal{K})\sin\vartheta\cos\Phi$ & $(-\mathcal{F},\mathcal{K})\sin\Phi$\tabularnewline
\hline
\multicolumn{2}{|c||}{$\gamma_{3}'$} & $(-\mathcal{F},\mathcal{K})\sin\Phi$ & $(\mathcal{F},-\mathcal{K})\sin\vartheta\cos\Phi$ & \textcolor{white}{$\dfrac{}{}$}$(0,0)$ \textcolor{white}{$\dfrac{}{}$} & $(\mathcal{F},-\mathcal{K})\cos\vartheta\cos\Phi $\tabularnewline
\hline
\multicolumn{2}{|c||}{$\gamma_{4}'$} & $(\mathcal{F},-\mathcal{K})\sin\vartheta\cos\Phi$ & $(\mathcal{F},-\mathcal{K})\sin\Phi$& $(\mathcal{F}, -\mathcal{K})\cos\vartheta\cos\Phi$ & \textcolor{white}{$\dfrac{}{}$}$(0,0)$ \textcolor{white}{$\dfrac{}{}$}
\tabularnewline
\hline
\end{tabular}

\label{tab:fusion-rule-two-island}
\end{table*}

\section{Fusion of MZMs\label{sec:Fusion-of-Majorana}}

The fusion of MZMs is mediated by hopping interaction which, according to the bulk Hamiltonian, corresponds to the operators $\hat{T}_{x}=ivs_{z}\sigma_{x}/2+m\tau_{z}\sigma_{z}$ and $\hat{T}_{y}=iv\tau_{z}\sigma_{y}/2+m\tau_{z}\sigma_{z}$ in $x$ and $y$ directions, respectively. We thus estimate the fusion strength between two MZMs, $\gamma_{i}$ and $\gamma_{j}'$ (with $i,j\in\{1,2,3,4\}$), by $\langle\Psi_{i}|\hat{T}_{x}|\Psi_{j}'\rangle$ and $\langle\Psi_{i}|\hat{T}_{y}|\Psi_{j}'\rangle$. Table\ \ref{tab:fusion-rule-two-island}
summarizes the results of the fusion between two sets of MZMs, $\{\gamma_{i}\}$ and $\{\gamma_{i}'\}$, which belong to a single or two connected SOTS islands. $2\Phi$ is the phase difference between the pairing potential in the two islands. $\varphi_{I}$ and $\varphi_{II}$ are the angle positions of the two MZMs with respect to the island from which they stem. For simplicity, we consider that the two islands have the same
model parameters except magnetic field direction and pairing phase. By taking $\{\gamma_{i}'\}=\{\gamma_{i}\}$, $\varphi_{I}=\varphi_{II}=\varphi$ and $\Phi=0$, the table corresponds to the results for a single island.

\section{Braiding properties \label{sec:Non-Abelian-Manipulations}}

\subsection{Exchange rule}

Let us first consider the adiabatic process of exchanging the two MZMs, $\gamma_{\alpha}$ and $\gamma_{\beta}$, in the triangle setup four times in succession by rotating the in-plane magnetic field from $\theta=0$ to $2\pi$. After this rotation, the system (Hamiltonian) returns to the initial configuration. Accordingly, the wavefunctions of $\gamma_{\alpha}$ and $\gamma_{\beta}$ recover their initial ones up to a Berry phase. The Berry phases of $\gamma_{\alpha}$ and $\gamma_{\beta}$ accumulated in the exchange period $[0,T]$ are given by
\begin{equation}
\Phi_{\nu}(T)=\int_{0}^{T}[\mathcal{A}(t)]_{\nu\nu}dt,\ \ \nu\in\{\alpha,\beta\}
\end{equation}
where $[\mathcal{A}(t)]_{\alpha\beta}=i\langle\psi_{\alpha}(t)|\partial_{t}|\psi_{\beta}(t)\rangle$ is the Berry connection matrix. Since the two MZMs are always well separated from each other, $\mathcal{A}(t)$ is diagonal. In numerical calculations, we discretize the period into a large number $N$ of equidistant intervals. At $t_{j}=j\Delta t$ with $\Delta t=T/N$ and $j\in\{1,...,N\}$, the direction of the magnetic field is $\theta_{j}=\pi j/N$ (with respect to the $x$ direction). When $\Delta t$ is very small such that the wavefunctions at neighboring steps are nearly the same, we approximate
\begin{equation}
e^{i[\mathcal{A}(t_{j})]_{\nu\nu}\Delta t}\approx\langle\psi_{\nu}(t_{j})|\psi_{\nu}(t_{j-1})\rangle\equiv F_{\nu}(t_{j}).\label{eq:approximation}
\end{equation}
Thus, we can calculate the Berry phases by \citep{Pahomi19arXiv}
\begin{equation}
\rho_{\nu N}e^{i\Phi_{\nu}} = \prod_{j=0}^{N}F_{\nu}(t_{j}).
\end{equation}

\begin{figure}[h]
\includegraphics[width=0.95\columnwidth]{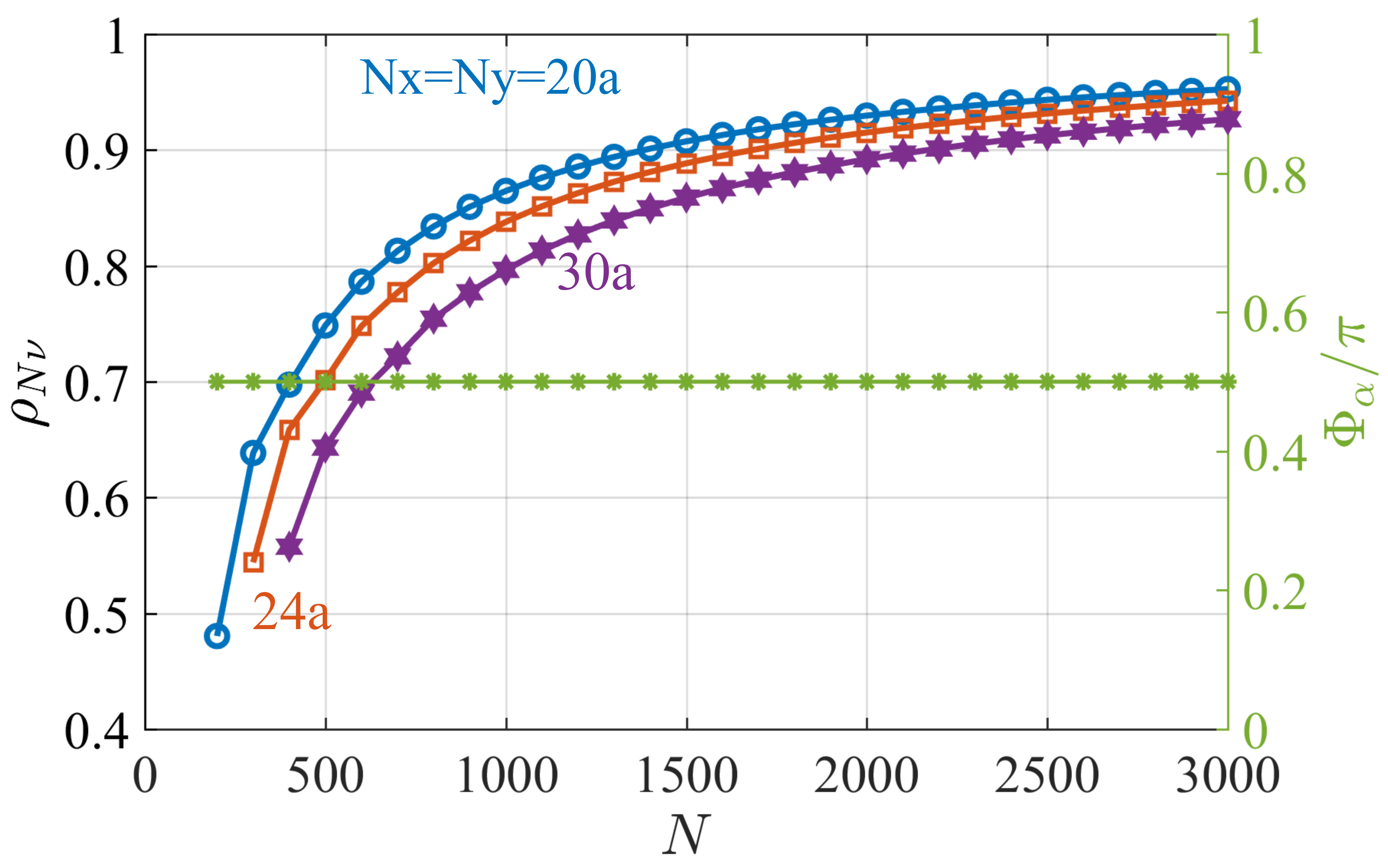}

\caption{The factor $\rho_{\nu N}$ and the Berry phase $\Phi_{\nu}$ of the first half of the period $[0,T/2]$ as functions of $N$ for the triangle of different sizes $N_{x}=N_{y}=20a$, $24a$ and $30a$, respectively. The same results are obtained in the second half of the period $[T/2,T]$.}

\label{fig:overlap_function-abs}
\end{figure}

According to our numerical results, the two MZMs have the same factor $\rho_{\nu N}$ and the same Berry phase $\Phi_{\nu}$. In Fig.\ \ref{fig:overlap_function-abs}, we plot $\rho_{\nu N}$ and $\Phi_{\nu}$ for increasing $N$. It shows that $\rho_{\nu N}$ depends on the number of steps $N$ and saturates to unity in the limit $N\rightarrow\infty$. For large $N$, each MZM accumulates a half quantized Berry phase $\pi/2$ during the first half of the period $[0,T/2]$ and another half quantized Berry phase $\pi/2$ during the second half of the period $[T/2,T]$. These half quantizated Berry phases are guaranteed by inversion symmetry of the minimal model. Thus, in a complete period, a quantized Berry phase $\pi$ is achieved. The relations between the wavefunctions of the MZMs in the final and  initial states are given by
\begin{equation}
\psi_{\nu}(t=T) = e^{i\Phi_{\nu}(T)}\psi_{\nu}(t=0)=-\psi_{\nu}(t=0).
\end{equation}
This implies that the four-time exchange operation transforms the MZMs as
\begin{equation}
\gamma_{\nu} \mapsto -\gamma_{\nu},\ \ \ \nu\in\{\alpha,\beta\}.\label{eq:Relation-four-times}
\end{equation}
Both MZMs flip sign after the operation. This feature can be exploited for topological quantum computation, which we explain below.

\begin{figure}[b]
\includegraphics[width=0.95\columnwidth]{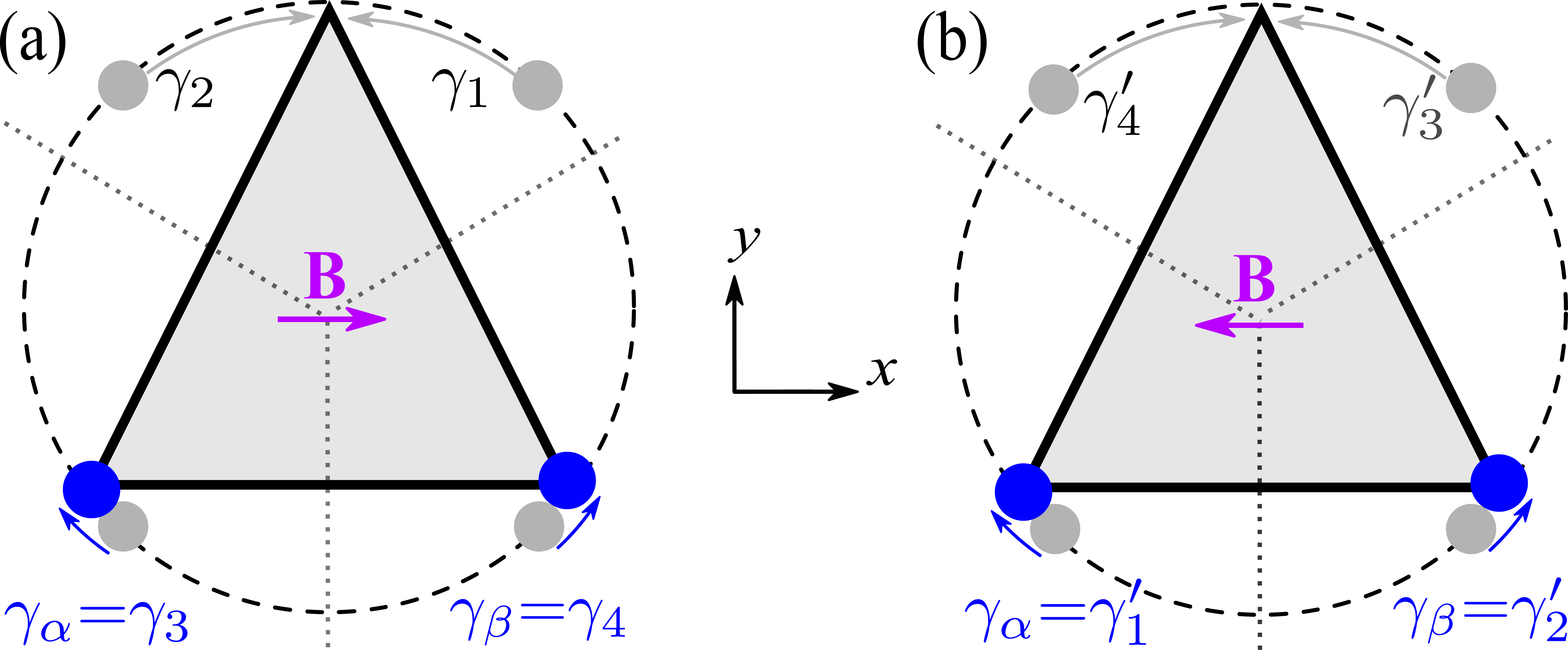}

\caption{Identification of $\gamma_{\alpha}$ and $\gamma_{\beta}$ with the MZMs $\{\gamma_{i}\}$ on a disk boundary (dashed circle)
when the magnetic field is in the (a) $\theta=0$ and (b) $\pi$ directions.}

\label{fig:mapping}
\end{figure}

We next consider the twice exchange of the two MZMs performed, for instance, in the first half of the period $[0,T/2]$ by rotating the magnetic field from $\theta=0$ to $\pi$. The wavefunctions of $\gamma_{\nu}$ at $t_f\equiv T/2$ can be written as
\begin{equation}
\psi_{\nu}(t=t_f)=e^{i\Phi_{\nu}(t_f)}\psi_{\nu}(\theta(t=t_f)),\label{eq:Relation}
\end{equation}
where $\psi_{\nu}(\theta(t=t_f))=\psi_{\nu}(\theta=\pi)$ are instantaneous eigenstates of the Hamiltonian $\mathcal{H}(\theta(t=t_f))$. To relate
$\psi_{\nu}(\theta=\pi)$ to the ones $\psi_{\nu}(\theta=0)$ in the initial state, it is instructive to identify $\gamma_{\alpha}$ and $\gamma_{\beta}$ with the four possible MZMs $\gamma_{i}$ ($i\in\{1,2,3,4\}$) on a disk geometry. As shown in Fig.\ \ref{fig:mapping},
$\gamma_{\alpha}$ corresponds to $\gamma_{3}$ in the initial state while it corresponds to $\gamma_{1}'$ in the final state: $\psi_{\alpha}(\theta=0)=\psi_{3}$ and $\psi_{\alpha}(\theta=\pi)=\psi_{1}'$. In this notation, the prime superscript indicates that the magnetic field applied in the triangle island is changed in the final state. Comparing $\psi_{1}'$ with $\psi_{3}$ [see Eqs.\ \eqref{psi1} and \eqref{psi3}] at the same corner, we find
\begin{eqnarray}
\psi_{1}' & = & \bar{\mathcal{P}}\psi_{3},\label{eq:other-part-of-BerryPhase}
\end{eqnarray}
where $\bar{\mathcal{P}}=\tau_zs_z$ is the inversion operator we introduced in Sec.\ \ref{sec:wavefunction} and the phase factors have been included in $\psi_{1}'$ ad $\psi_{3}$ so that they are written on the same basis. Plugging Eq.\ (\ref{eq:other-part-of-BerryPhase}) in Eq.\ (\ref{eq:Relation}), we obtain
\begin{equation}
\psi_{\alpha}(t=t_f)= e^{i\pi/2} \bar{\mathcal{P}} \psi_{\alpha}(t=0).\label{eq:double-exchange_WF_a}
\end{equation}
In a similar way, we find
\begin{equation}
\psi_{\beta}(t=t_f)= e^{i\pi/2} \bar{\mathcal{P}} \psi_{\beta}(t=0).\label{eq:double-exchange_WF_b}
\end{equation}
Equations\ (\ref{eq:double-exchange_WF_a}) and (\ref{eq:double-exchange_WF_b}) imply that the double exchange transforms the MZMs as:
\begin{equation}
\gamma_{\nu}\mapsto e^{i\pi/2}\bar{\mathcal{P}} \gamma_{\nu},\ \ \nu\in\{\alpha,\beta\}.
\label{eq:flip-Majorana_a}
\end{equation}
Different from the usual Majorana exchange, the MZMs are transformed to their inversion symmetry counterparts whose wavefunctions are connected to the original ones by the transformation $\bar{\mathcal{P}}$. Exchanging $\gamma_\alpha$ and $\gamma_\beta$ twice again, we obtain the result in Eq.\ \eqref{eq:Relation-four-times}.

\begin{figure}[h]
\includegraphics[width=0.95\columnwidth]{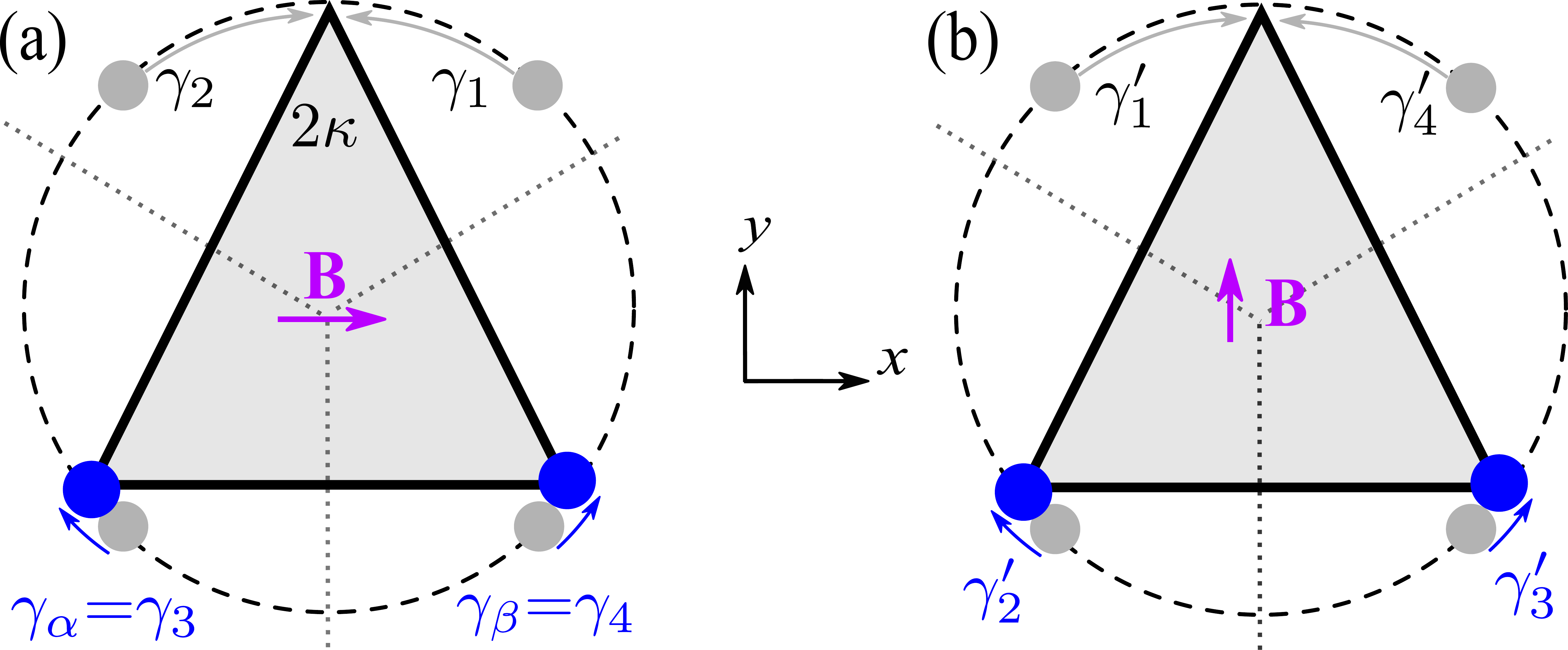}

\caption{Identification of $\gamma_{\alpha}$ and $\gamma_{\beta}$ with the MZMs $\{\gamma_{i}\}$ on a disk boundary when the magnetic field is in the (a) $\theta=0$ and (b) $\pi/2$ directions.}

\label{fig:mapping-one-exchange}
\end{figure}

Now, we look at the single exchange of $\gamma_{\alpha}$ and $\gamma_{\beta}$ by rotating the magnetic field from $\theta=0$ to $\pi/2$. Again we relate the instantaneous eigenstates $\psi_{\nu}(\theta=\pi/2)$ to $\psi_{\nu}(\theta=0)$ with the help of the auxiliary disk geometry, see Fig.\ \ref{fig:mapping-one-exchange}. In the initial state, $\gamma_{\alpha}$ and $\gamma_{\beta}$ corresponds to $\gamma_{3}$ and $\gamma_{4}$,
respectively, while they corresponds to $\gamma_{2}'$ and $\gamma_{3}$ in the final state, respectively. Namely,
\begin{align}
\psi_{\alpha}(\theta=0) & =\psi_{3},\ \ \psi_{\alpha}(\theta=\pi/2)=\psi_{2}',\nonumber \\
\psi_{\beta}(\theta=0) & =\psi_{4},\ \ \psi_{\beta}(\theta=\pi/2)=\psi_{3}'.
\end{align}
Comparing the wavefunctions $\psi_{2}'$ with $\psi_{4}$, we find
\begin{eqnarray}
\psi_{2}' & = &\mathcal{T}_{2\pi-4\kappa}\tau_zs_z \psi_{4},
\end{eqnarray}
where the matrix $\mathcal{T}_{\chi}=e^{i\chi/2}e^{-i\chi\tau_{z}s_{z}\sigma_{z}/2}$ stems from different positions of $\gamma_{2}'$ and $\gamma_{4}$ on the disk boundary and $2\kappa=2\text{arctan}(N_{x}/2N_{y})$ is the angle of the upper corner, see Fig.\ \ref{fig:mapping-one-exchange}(a). Thus, we obtain
\begin{equation}
\psi_{\alpha}(t=t_f)=e^{i\Phi_{\alpha}(t_f)} e^{-2i\kappa} \mathcal{T}_{-4\kappa}\tau_zs_z\psi_{\beta}(t=0) \label{eq:exchange_aa}
\end{equation}
with $t_f=T/4$. Comparing $\psi_{3}'$ with $\psi_{3}$, we find
\begin{equation}
\psi_{3}' = e^{2i\kappa}\mathcal{T}_{4\kappa}\psi_{3}.\label{eq:exchange_b}
\end{equation}
Hence,
\begin{eqnarray}
\psi_{\beta}(t =t_f)=e^{i\Phi_{\beta}(t_f)} e^{2i\kappa}\mathcal{T}_{4\kappa}\psi_{\alpha}(t=0).\label{eq:exchange_bb}
\end{eqnarray}
We can rewrite Eqs.\ (\ref{eq:exchange_aa}) and (\ref{eq:exchange_bb}) as
\begin{eqnarray}
\psi_{\alpha}(t=t_f) & = & \mathcal{S}\mathcal{\mathcal{D}}\psi_{\beta}(t=0),\nonumber \\
\psi_{\beta}(t=t_f) & = & \mathcal{S} \psi_{\alpha}(t=0),\label{eq:final-exchange}
\end{eqnarray}
where $\mathcal{S} = e^{i\Phi_{\beta}(t_f)} e^{2i\kappa}\mathcal{T}_{4\kappa}$, $\mathcal{\mathcal{D}}=e^{-4i\kappa} \mathcal{T}_{-8\kappa} \bar{\mathcal{P}}$ and we have used the numerical result $\Phi_{\beta}(t)=\Phi_{\alpha}(t)$.
The two MZMs evolve in different ways during the exchange process. The presence of two MZMs yields two degenerate ground states with different fermion parity in the system. The different evolutions of the MZMs indicate that the two ground states are transformed in different ways.

\subsection{Non-Ablelian quantum gates}

To elucidate the non-Abelian property of MZM braiding and exploit them for quantum computation, we now consider four MZMs: $\gamma_{a}$, $\gamma_{b}$, $\gamma_{c}$ and $\gamma_{d}$, and braiding them four times in each performance. We may define two complex fermions as
\begin{align}
f_{bc}=(\gamma_{b}+i\gamma_{c})/2,\ \ f_{ad}=(\gamma_{d}+i\gamma_{a})/2.
\end{align}
Given a fixed global fermion parity. There are two degenerate ground states of the system, forming the qubit for computation. Without loss of generality, we assume these qubit states as $\{|00\rangle,|11\rangle\}$, where $n_{bc}\in\{0,1\}$ in the Dirac notation $|{n_{bc}n_{ad}}\rangle$ indicates the absence and presence of $f_{bc}$ and $n_{ad}\in\{0,1\}$ indicates the absence and presence of $f_{ad}$.
According to Eq.\ \eqref{eq:Relation-four-times}, the operator for braiding $\gamma_a$ and $\gamma_b$ four times can be formulated as
\begin{eqnarray}
  \hat{T}_{ab}^{(4)}&=&\gamma_a\gamma_b = i(f^\dagger_{ad}-f_{ad})(f_{bc}^\dagger+f_{bc}). \label{eq:exchange_ab}
\end{eqnarray}
Acting $\hat{T}_{ab}^{(4)}$ on the qubit, we find
\begin{eqnarray}
\hat{T}_{ab}\begin{pmatrix}|{00}\rangle\\
|{11}\rangle
\end{pmatrix} & = & -i \begin{pmatrix}0 & 1\\
1 & 0
\end{pmatrix}\begin{pmatrix}|{00}\rangle\\
|{11}\rangle
\end{pmatrix}.
\end{eqnarray}
It acts like a $\sigma_x$ gate on the qubit and switches the qubit.
For the four-time braiding of $\gamma_c$ and $\gamma_d$ with the operator:
\begin{eqnarray}
  \hat{T}_{cd}^{(4)}&=&\gamma_c\gamma_d = i(f_{bc}^\dagger-f_{bc})(f^\dagger_{ad}+f_{ad}), \label{eq:exchange_ab}
\end{eqnarray}
we find similarly
\begin{eqnarray}
\hat{T}_{cd}\begin{pmatrix}|{00}\rangle\\
|{11}\rangle
\end{pmatrix} & = & i \begin{pmatrix}0 & 1\\
1 & 0
\end{pmatrix}\begin{pmatrix}|{00}\rangle\\
|{11}\rangle
\end{pmatrix}.
\end{eqnarray}
It is also a $\sigma_x$ gate for the qubit.
Finally, we can see that the four-time exchange of $\gamma_b$ and $\gamma_c$ with the operator:
\begin{eqnarray}
  \hat{T}_{bc}^{(4)}&=&\gamma_b\gamma_c = i(f_{bc}^\dagger-f_{bc})(f^\dagger_{ad}+f_{ad}) \label{eq:exchange_ab}
\end{eqnarray}
plays as a $\sigma_z$ gate on the qubit, i.e.,
\begin{eqnarray}
\hat{T}_{cd}\begin{pmatrix}|{00}\rangle\\
|{11}\rangle
\end{pmatrix} & = & i \begin{pmatrix}1 & 0 \\
0 & -1
\end{pmatrix}\begin{pmatrix}|{00}\rangle\\
|{11}\rangle
\end{pmatrix}.
\end{eqnarray}
These quantum gates are clearly non-commutative.

\section{Basic operations of braiding Majorana zero modes \label{sec:basic-braiding}}

\subsection{Rotating magnetic fields}
In this section, we discuss the operations of the two exchanges $\gamma_{b}\leftrightarrow\gamma_{c}$ and $\gamma_{c}\leftrightarrow\gamma_{d}$ in the trijunction setup. The two exchanges, together with the one $\gamma_{a}\leftrightarrow\gamma_{b}$ that we have illustrated in Sec.\ \ref{sec:braiding-more-modes} in the main text, can generate the whole braid group of the four MZMs $\gamma_{a}$, $\gamma_{b}$, $\gamma_{c}$, and $\gamma_{d}$.

To exchange $\gamma_b$ and $\gamma_c$, we can simply rotate the magnetic field $\bm{B}_2$ in $T2$. Every time we increase the direction $\theta_2$ of $\bm{B}_2$ by $\pi/2$, we exchange $\gamma_b$ and $\gamma_c$ once as we have shown in the single triangle setup.

The operation for exchanging $\gamma_c$ and $\gamma_d$ is similar to that for the exchange $\gamma_{a}\leftrightarrow\gamma_{b}$ introduced in Sec.\ \ref{sec:braiding-more-modes}. Explicitly, we do as follows: (1) turn $\theta_2=-\pi/3\rightarrow-\pi/6$; (2) move $\gamma_c$ to the center; (3) increase $\theta_3$ by $\pi/2$ which exchanges the positions of $\gamma_c$  and $\gamma_d$ in the island $T3$; (4) turn back $\theta_2=-\pi/6\rightarrow-\pi/3$; and (5) move $\gamma_d$ to the position where $\gamma_c$ was located initially.

\subsection{Tuning directions and strengths of magnetic fields}
We next present the operations for the exchanges $\gamma_b\leftrightarrow\gamma_c$ and $\gamma_c\leftrightarrow\gamma_d$ using the T-junction protocol.

The operations for exchanging $\gamma_b$ and $\gamma_c$ can be performed as follows: (1) turn $\theta_2=-\pi/3\rightarrow-\pi/2$; (2) turn $\theta_3=-\pi/3$ to $-\pi/6$; (3) decrease $B_2$ from $B_>$ to $B_<$; (4) turn $\theta_1=-\pi/3\rightarrow -\pi/2$; (5) turn $\theta_3=-\pi/6$ to $-\pi/3$; (6) increase $B_2$ back to $B_>$; (7) turn $\theta_1=-\pi/2\rightarrow -\pi/3$; and (8) turn $\theta_2=-\pi/2\rightarrow -\pi/3$.

The operations for exchanging $\gamma_c$ and $\gamma_d$ can be performed as follows: (1) turn $\theta_2=-\pi/3\rightarrow-\pi/6$; (2) turn $\theta_1=-\pi/3$ to $-\pi/2$; (3) decrease $B_3$ from $B_>$ to $B_<$; (4) turn $\theta_2=-\pi/6\rightarrow -\pi/3$; (5) turn $\theta_1=-\pi/2$ to $-\pi/3$; and (6) increase $B_3$ back to $B_>$.

\end{document}